\documentclass[12pt,fleqn]{article}

\usepackage{amsmath,amsfonts,amssymb}
\usepackage{latexsym}
\usepackage{amsmath} 
\usepackage{amssymb} 
\usepackage{bm}
\usepackage{graphicx}
\usepackage[numbers, longnamesfirst,nonamebreak]{natbib}

\newcommand{\bra}{\left\langle}
\newcommand{\ket}{\right\rangle}
\newcommand{\ld}{\mathcal{L}}

\newcommand{\pd}[2]{\frac{\partial #1}{\partial #2}}

\newcommand{\hide}[1]{}

\newcommand{\Cn}[1]{\begin{center} #1 \end{center}}

\newcommand{\be}{\begin{equation}}
\newcommand{\ee}{\end{equation}}

\newcommand{\heading}[1]{\begin{center} \Large {#1} \end{center}}

\makeatletter
\newcommand\ackname{Acknowledgements}
\if@titlepage
  \newenvironment{acknowledgements}{%
      \titlepage
      \null\vfil
      \@beginparpenalty\@lowpenalty
      \begin{center}%
        \bfseries \ackname
        \@endparpenalty\@M
      \end{center}}%
     {\par\vfil\null\endtitlepage}
\else
  \newenvironment{acknowledgements}{%
      \if@twocolumn
        \section*{\abstractname}%
      \else
        \small
        \begin{center}%
          {\bfseries \ackname\vspace{-.5em}\vspace{\z@}}%
        \end{center}%
        \quotation
      \fi}
      {\if@twocolumn\else\endquotation\fi}
\fi
\makeatother

\begin{document} 
 \normalsize
\begin{titlepage}

\heading{Monopole Solutions in a Two Measure Field Theory and the Evolution of Vacuum Bubbles}

\bigskip
\bigskip

\Cn{\em Thesis submitted in partial fulfillment of the requirement for the degree of master of science in the Faculty of Natural Sciences}

\bigskip
\bigskip

\Cn{

Submitted by: {\bf Alon Yaniv}

\bigskip

Advisor: {\bf Prof. Eduardo Guendelman} }

\bigskip
\bigskip

\Cn{ Department of Physics}
\Cn{ Faculty of Natural Sciences}
\Cn{ Ben-Gurion University of the Negev}

\bigskip
\bigskip
\begin{center}
\today
\end{center}
\bigskip
\bigskip
\bigskip
\bigskip

 \begin{center}
 \begin{tabular}{llrl}
Author's Signature & \_\_\_\_\_\_\_\_\_\_\_\_\_\_\_\_\_\_ &\ \ \ \ \  Date: & \_\_\_\_\_\_\_\_\_\_\_\_\_\_\_\_\_\_ \cr
Supervisor's Approval  & \_\_\_\_\_\_\_\_\_\_\_\_\_\_\_\_\_\_ & Date: & \_\_\_\_\_\_\_\_\_\_\_\_\_\_\_\_\_\_\cr
Faculty Council Approval & \_\_\_\_\_\_\_\_\_\_\_\_\_\_\_\_\_\_ &  Date: & \_\_\_\_\_\_\_\_\_\_\_\_\_\_\_\_\_\_
 \end{tabular}
 \end{center}
 
\end{titlepage}

\numberwithin{equation}{section}

\begin{abstract}
The existence of point-like topological defects, or monopoles, is an inevitable consequence of many grand unified theories(GUTs)\cite{Weinberg1982445}\cite{AlexanderVilenkin2000}\cite{Hooft1974}\cite{Preskill1984}. This led to extensive research of these objects starting from the late 60's\cite{Hooft1974}\cite{Wu1968}\cite{Polyakov1974}\cite{Hooft1976}\cite{Coleman1983}. In recent years as experiments failed to provide evidence\cite{Milton:2006cp} of the existence of these models research has somewhat shifted away. One key line of research was in determining the gravitational field generated in the vicinity of monopoles \cite{Cho1975}\cite{Bais1975}\cite{Barriola1989}\cite{Gibbons} and determining possible connections between monopoles and cosmological scenarios \cite{Preskill1979}\cite{Guendelman1991a}\cite{Nucamendi2001}\cite{Sakai:2006fg}.

Two Measure Field Theory(TMT) is a generally coordinate invariant theory in which the main supposition  is that for describing the effective action for 'gravity $+$ matter' at energies below the Planck scale, the usual form of the action $S = \int L\sqrt{-g}d^{4}x$ is not complete. We hypothesize that the effective action
has to be of the form \cite{Guendelman:1998ms}\cite{Guendelman:2004jm}\cite{Guendelman:2004bn}\cite{Guendelman2006c}\cite{Guendelman2007}\cite{Guendelman2008k}\cite{Guendelman2010f}
\begin{equation}\nonumber
    S = \int L_{1}\Phi d^{4}x +\int L_{2}\sqrt{-g}d^{4}x,
\label{S}
\end{equation}
 including two Lagrangians $ L_{1}$ and $L_{2}$ and two measures of integration $\sqrt{-g}$ and $\Phi$. 

	 We adopted the framework of TMT for the study of magnetic monopoles and discovered some interesting features which are absent from the classical theory. Our calculations focused on the regions far away from the monopole core, which are usually the regions of interest. We found that the gravitational field exhibits both a Reissner-Nordstr$\ddot{\text{o}}$m type behavior predicted in theories where the monopole was produced by a local symmetry breaking\cite{Cho1975}\cite{Bais1975}, and a deficit angle predicted in theories with global symmetry breaking\cite{Barriola1989}\cite{Guendelman1991a}.
 
 Next, following the footprints of Guendelman and Rabinowitz \cite{Guendelman1991a}, we wanted to estimate the effect these solutions might have if they are to surround a vacuum bubble. We adapt the technique of using Israel's junction conditions\cite{Israel1966} derived by Blau et al.\cite{Blau1987} to study the evolution of vacuum bubbles. We found new types of behavior for the bubble dynamics which are absent from known classical solutions. We suggest also a classical mechanism in which our solution might undergo a phase transition which will cause the initially stable bubble to undergo inflation. Another interesting possibility, in light of work done by Kawai and Matsuo\cite{Kawai2010}, is the existence of classical stable particle-like solutions. These are stable compact configurations which for an outside observer seems to be characterized only by their mass and their charge.
\end{abstract}
\clearpage
\begin{acknowledgements}
I owe my deepest gratitude to my advisor, Prof. Eduardo Guendelman, for infinite patience and guidance. I would like to thank also Dr. Alexander Kaganovich for very helpful and enlightening conversations.\\

Lastly, this thesis would not have been possible without the support of my family and friends to which I am forever grateful.\\

\textit{Alon Yaniv}
\end{acknowledgements}
\clearpage
\tableofcontents
\clearpage
\listoffigures
\clearpage
\section{Introduction to Spontaneous symmetry breaking}\label{sec:introSSB}
Spontaneous symmetry breaking (SSB) as an idea originated in condensed matter physics. A familiar example is the isotropic model of a ferromagnet which although described by a rotationally symmetric hamiltonian, can develop a net magnetic moment pointing in some direction. In modern theories of elementary particles, symmetry breaking is described in terms of scalar fields, usually called Higgs fields. The characteristic trait of a spontaneously broken theory is that in the ground state, the Higgs field acquires a non-zero expectation value (vacuum expectation value or VEV) which leads to a theory which does not exhibit all the symmetries of the hamiltonian.
\footnote{Throughout we shall use a metric with signature (+,-,-,-) unless stated otherwise. Einsteins field equations are used in the form $G_{\mu\nu}=\frac{\kappa}{2}T_{\mu\nu}$, where $\kappa=16\pi G$ and $G$ is Newtons constant. The Riemann curvature tensor is given by $R^\lambda_{\mu\nu\sigma}(\Gamma)=\Gamma^\lambda_{\mu\nu,\sigma}-\Gamma^\lambda_{\mu\sigma,\nu}+\Gamma^{\lambda}_{\alpha\sigma}\Gamma^{\alpha}_{\mu\nu}-\Gamma^{\lambda}_{\alpha\nu}\Gamma^{\alpha}_{\mu\sigma}$.}
\subsection{Global Abelian Symmetry Breaking}

\begin{figure}[h]
\begin{center}
\includegraphics[width=3.3in]{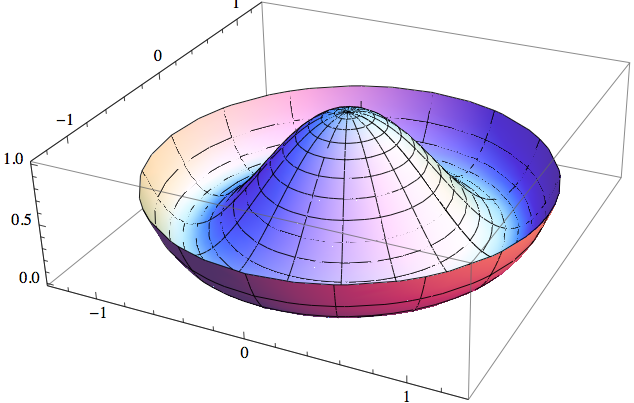}
\caption{\label{Fig.mexicanhat} The mexican hat potential $ V(\phi)=\frac{1}{4}\lambda\left({\phi^\dagger}\phi-\eta^2\right)^2
$ . The field takes its vacuum expectation value at the global minima $|\phi|=\eta$.}
\end{center}
\end{figure}

	The essential features of SSB can be illustrated using a simple model first studied by Goldstone\cite{Goldstone1961}. This has the classical Lagrangian density 
\be\label{eq:goldstonelagrangian}
\ld=\partial_{\mu}{\phi^\dagger}\partial^{\mu}{\phi}-V(\phi),
\ee
with $\phi$ a complex scalar field and the potential $V(\phi)$ taken to be the mexican hat potential  given by
 \be\label{eq:mexicanhat}
 V(\phi)=\frac{1}{4}\lambda\left({\phi^\dagger}\phi-\eta^2\right)^2,
 \ee
 with positive constants $\lambda$ and $\eta^2$. This potential is depicted in figure (\ref{Fig.mexicanhat}). The model is symmetric under $U(1)_{\text{global}}$ transformations 
\be\label{eq:globalphase}
\phi(x)\rightarrow e^{i\alpha}\phi(x),
\ee
where 'global' indicates that the phase $\alpha$ is independent of $x$. The minima of the potential lies on the circle $|\phi|=\eta$ so the vacuum of the theory is characterized by a non-zero VEV $\bra0|\phi|0\ket=\eta e^{i\theta}$ where $\theta$ is an arbitrary phase. The phase transformation (\ref{eq:globalphase}) changes $\theta$ to $\theta+\alpha$ hence the vacuum state $|0\rangle$ is not invariant under (\ref{eq:globalphase}) and the symmetry is spontaneously broken.

The state with $\bra0|\phi|0\ket=0$ corresponds to a local maximum of $V(\phi)$. We see this by observing that the squared mass which is proportional to the frequency of small oscillations $m^2\propto\left[\pd{}{\phi}\pd{}{\bar{\phi}}V(\phi)\right]_{\phi=0}=-\frac{1}{2}\lambda\eta^2$ is negative. The negative squared mass indicates the instability of the symmetric state.

The broken symmetry vacua with different phase $\theta$ are all equivalent so without loss of generality we can study any one of them. Choosing $\theta=0$ we can represent $\phi$ as
\be
\phi=\eta+\frac{1}{\sqrt{2}}\left(\phi_1+i\phi_2\right).
\ee
 $\phi_1$ and $\phi_2$ are two real fields with zero VEV. Substituting this into the Lagrangian density (\ref{eq:goldstonelagrangian}) gives
 \be
 \ld=\frac{1}{2}\left(\partial_{\mu}\phi_1\right)^2+\frac{1}{2}\left(\partial_{\mu}\phi_2\right)^2-\frac{1}{2}\lambda\eta^2\phi_1^2+\ld_{\text{int}},
 \ee
where the interaction term includes terms of higher order of the fields. We see that $\phi_1$ represents a particle with positive mass $\mu=\sqrt{\lambda}\eta$ while $\phi_2$ is massless. The intuitive description of this result is that $\phi_1$ corresponds to oscillations about a point on the circle of minima $\phi=\eta$ while $\phi_2$ corresponds to motion around the circle. The appearance of massless scalar particles, called Goldstone bosons, is a general feature of spontaneously broken global symmetries.

 \subsection{Local Abelian Symmetry Breaking - The Higgs Model}
 Although global symmetries are of considerable interest, the central role in particle physics is played by gauge theories with spontaneously broken symmetries. The simplest example, known as the abelian Higgs model \cite{Higgs1964}, describes scalar electrodynamics with Lagrangian density
 \be\label{eq:higgslag}
 \ld=\bar{D}_{\mu}\bar{\phi}D^{\mu}\phi-\frac{1}{4}F_{\mu\nu}F^{\mu\nu}-V(\phi).
 \ee
$\phi$ again is a complex scalar field and the covariant derivative is given by $D_{\mu}=\partial_{\mu}-ieA_{\mu}$. The field tensor $F_{\mu\nu}$ is given by $F_{\mu\nu}=\partial_{\mu}A_{\nu}-\partial_{\nu}A_{\mu}$ where $A_{\mu}$ is a gauge vector field, $e$ - a gauge coupling and the potential $V(\phi)$ is given by (\ref{eq:mexicanhat}). This model is invariant under $U(1)_{\text{local}}$ with the transformation
\be\label{eq:localphase}
\phi(x)\rightarrow e^{i\alpha(x)}\phi(x),\qquad A_{\mu}(x)\rightarrow A_{\mu}(x)+\frac{1}{e}\partial_{\mu}\alpha(x).
\ee
Since $V(\phi)$ has its minima at $|\phi|=\eta$, this symmetry is spontaneously broken and the field acquires a non zero VEV. To study the properties of the new vacuum, it is convenient to use the gauge in which $\phi(x)$ is real. Then, representing $\phi$ as $\phi=\eta+\frac{1}{\sqrt {2}}\phi_1$ we obtain 
\be
\ld=\frac{1}{2}\left(\partial_{\mu}\phi_1\right)^2-\frac{1}{2}\mu^2\phi_1^2-\frac{1}{4}F_{\mu\nu}F^{\mu\nu}+\frac{1}{2}M^2A_\mu A^\mu+\ld_{\text{int}},
\ee
 where
 \be
 \mu=\sqrt{\lambda}\eta,\qquad M=\sqrt{2}e\eta,
 \ee
 and the interaction term includes higher order terms of both $\phi$ and $A_{\mu}$. We see that now there are no massless Goldstone bosons. Instead, the corresponding degree of freedom has been absorbed into the vector field, making it massive and therefore allowing for a third independent polarization as opposed to the original two.
 
  \subsection{Global Non-Abelian Symmetry Breaking}
The simple goldstone model (\ref{eq:goldstonelagrangian}) can be generalized to possess an invariance under an arbitrary group $G$ of global gauge transformations. These transformation are realized on an $n$-component scalar field $\phi_i$ through the action of the matrix representation of the group $G$,
\be\label{eq:gentrans}
\phi^{i}\rightarrow \phi_{i}'=D_{ij}(g)\phi_{j},
\ee 
 where $D(g)$ is an $n\times n$ matrix and $g\in G$. The representation is a mapping from $G$ to the operators $D$ which act on the vector space spanned by the fields $\phi_i$ and preserve group multiplication $D(g_1)D(g_2)=D(g_1g_2)$. Elements of the Lie group $\text{G}$ of dimension $N$ can be written in the form 
 \be
 g=\exp(-i\omega_aL^a),
 \ee
 where $L^a$ are the $N$ group generators and $\omega_a$ are some real arbitrary numbers. The generators form the lie algebra of $G$ and satisfy the commutation relation
 \be\label{eq:liealgebra}
 \left[L^a,L^b\right]=-if_{abc}L^{c}, 
 \ee
 where the $f_{abc}$'s are the structure constants of $G$. A matrix representation of any dimension can be generated by finding a set of $n\times n$ matrices $T^a$ which satisfy the same commutation relations. In-fact, $G$ itself can be thought of as being a group of matrices by taking any faithful representation of $G$ with Hermitian generators ,$L^a$, as the basis for the Lie algebra of $G$. Another example is the N-dimensional representation generated by the structure constants themselves, $(T^a)_{bc}=-if_{abc}$, which is known as the adjoint representation.  For a compact Lie group $\text{G}$, we can normalize the generators by 
 \be\label{eq:norma}
 \text{Tr}\{T^{a}T^{b}\}=\delta_{ab},
 \ee
 and we shall assume the correspondence, $T^a=D(L^a)$ between the two sets of generators.
 With this in mind we can consider the generalized Lagrangian with the choice $\varphi=\{\phi_i\}$ to be $n$ real scalar fields in a real unitary representation of $G$ with generators $T^a$
 \be\label{eq:genlag}
 \ld=\frac{1}{2}(\partial_\mu \varphi^\dagger)(\partial^\mu \varphi)-V(\varphi).
 \ee
 
Models with complex fields can be brought to this form by representing each of them by two real fields. The Lagrangian (\ref{eq:genlag}) is invariant under (\ref{eq:gentrans}) if $V\left(D(g)\varphi\right)=V(\varphi)$, which can be easily re-expressed as
\be\label{eq:invarcondition}
\pd{V}{\phi_i}T^{a}_{ij}\phi_{j}=0.
\ee
 If the minima of $V(\varphi)$ are at non-zero values of $\phi_i$, then the symmetry will be spontaneously broken, and the fields $\phi_i$ will acquire non zero VEVs $\bra0|\varphi|0\ket=\varphi_0$. The elements of $G$ which leave $\varphi_0$ unchanged form a group $H$ called the unbroken subgroup of $G$ with respect to $\varphi_0$. In terms of the matrices $D(g)$ in the representation of $G$, $H$ is defined by
 \be
 H=\{g\in G|D(g)\varphi_0=\varphi_0\}.
 \ee
 The generators $t^a$ of $H$ must satisfy $t^a\varphi_0=0$, we are free to choose the generators $T^a$ of the symmetry group $G$ such that $t^a$, the generators of
the unbroken subgroup $H$ are a subset of $T^a$. The generators $t^a$ are referred to as the unbroken generators of $G$ with the remainder of the generators $T^a$ being the broken ones.
Next, we can represent $\varphi$ as $\varphi=\varphi_0+\varphi'$ and expand the potential in powers of $\phi_i$. We find that as any first derivative of the potential vanishes in its minima,  small perturbations about $\varphi_0$ are described by the Lagrangian
\be
\ld=\frac{1}{2}(\partial_\mu \phi_i')(\partial^\mu \phi_i')-\frac{1}{2}\mu_{ij}^2\phi_i'\phi_j',
\ee
The mass matrix is given by
\be\label{eq:massmat}
\mu^2_{ij}=\left[\pd{}{\phi_i}\pd{}{{\phi}_j}V(\phi)\right]_{\phi=\phi_0},
\ee
and has non-negative eigenvalues. Differentiating (\ref{eq:invarcondition}) and using (\ref{eq:massmat}), we obtain
\be
\mu^2_{ij}T^a_{jk}\phi_{0k}=0,
\ee
but since all vectors $T^a\varphi_0$ formed from broken generators are linearly independent we arrive to the conclusion that the mass matrix must have a zero eigenvalue for any broken generator. These massless fields correspond to Goldstone bosons while the remaining masses(for the unbroken generators) are, in general, non-zero.
We shall note now that as the vacuum $\varphi_0$ was chosen arbitrarily, in general the expectation value of $\varphi$ in any other vacuum(imposed by symmetry) are of the form $D(g)\varphi_0$ for some $g\in G$. We can therefore identify a manifold of equivalent vacua $\mathcal{M}$ with the coset space
\be
\mathcal{M}=G/H,
\ee
which we shall call the vacuum manifold. Later we shall see that the topology of $\mathcal{M}$ has a crucial role in determining the nature of monopoles, as well as other topological defects.

  \subsection{Local Non-Abelian Symmetry Breaking - The Yang-Mills Model}
  
  Turning to the case of local gauge invariance, we introduce gauge fields $A_\mu^a$ which are associated with matrices in the Lie algebra of $G$ (\ref{eq:liealgebra}) by $\mathcal{A}_{\mu}=A_{\mu}^aL^a$. The non-Abelian generalization of the Higgs model (\ref{eq:higgslag}), first discussed by Yang and Mills\cite{Yang1954}, is then  
 \be\label{eq:nonabhiggs}
 \ld=\frac{1}{2}{D}_{\mu}{\varphi_i}D^{\mu}\varphi_i-\frac{1}{4}F^a_{\mu\nu}F^{a\mu\nu}-V(\varphi).
 \ee
where
\be
F^a_{\mu\nu}=\partial_{\mu}A_{\nu}^a-\partial_{\nu}A_{\mu}^a+ef_{abc}A_\mu^bA_\nu^c
\ee
is the Yang-Mills field strength, $A^a_{\mu}$ is a gauge vector field, $e$ is the gauge coupling constant, and ${D}_{\mu}{\varphi}$ is the gauge covariant derivative of $\varphi$
\be
{D}_{\mu}{\varphi}=\left(\partial_{\mu}-ieA^a_{\mu}T^a\right)\varphi.
\ee

The gauge transformations $g=g(x)$, which are now allowed to vary as functions of space-time coordinates, are defined by
\be\label{eq:symtransym}
\varphi\rightarrow D(g)\varphi,\qquad \mathcal{A}_{\mu}\rightarrow g\mathcal{A}_{\mu}g^{-1}+i\frac{1}{e}g^{-1}\partial_{\mu}g.
\ee
The Covariant derivative ${D}_{\mu}{\varphi}$ and the field strength $\mathcal{F}_{\mu\nu}=F^a_{\mu\nu}L^a$ have simple transformation properties
\be
{D}_{\mu}{\varphi}\rightarrow D(g){D}_{\mu}{\varphi}, \qquad \mathcal{F}_{\mu\nu}\rightarrow g\mathcal{F}_{\mu\nu}g^{-1}.
\ee
Using (\ref{eq:norma}) the last term in (\ref{eq:nonabhiggs}) can be written as $-\text{Tr}\{\mathcal{F}_{\mu\nu}\mathcal{F}^{\mu\nu}\}/4$.

When $\varphi$ develops a non-zero VEV $\varphi_0$, the symmetry (\ref{eq:symtransym})  is spontaneously broken. The properties of the broken symmetry state are most easily understood in the gauge in which the vector $\varphi_i$ has vanishing components in the subspace $\mathcal{G}$ defined by $T^a\varphi_0$. These fields would have corresponded to Goldstone bosons had the symmetry been  global. The dimension of $\mathcal{G}$ is equal to the number of broken generators(and the number of gauge choices one is allowed to make). Using again the substitution $\varphi=\varphi_0+\varphi'$, we get the Lagrangian describing the excitations above the broken symmetry vacuum
\be
\ld=\frac{1}{2}(\partial_\mu \phi_i')^2-\frac{1}{2}\mu_{ij}^2\phi_i'\phi_j'-\frac{1}{4}(F_{\mu\nu}^a)^2+\frac{1}{2}M_{ab}^2A_{\mu}^{a}A^{b\mu}+\ld_{\text{int}}.
\ee
The scalar field mass matrix is given by (\ref{eq:massmat}) with the added constraint that the summation over $i,j$ does not include components in the subspace $\mathcal{G}$. The vector field mass matrix is given by
\be
M^2_{ab}=e^2\left(T^aT^b\right)_{ij}\phi_{0i}\phi_{0j}
\ee

We see that the vector fields associated with broken generators acquire non-zero masses, while the gauge fields associated with the unbroken subgroup $H$ remain massless. The would-be goldstone bosons have disappeared and instead, the corresponding degrees of freedom have been absorbed into additional spin states of the massive vector fields. 
\clearpage
\section{Some Remarks on Homotopy Groups}

	When trying to determine whether defects appear at a particular symmetry breaking we have to look at the topology of the vacuum manifold $\mathcal{M}$. Domain walls can form if $\mathcal{M}$ has disconnected components, strings can form if $\mathcal{M}$ is not simply connected (contains unshrinkable loops), and for our purposes, monopoles can form if $\mathcal{M}$ contains unshrinkable surfaces. Luckily, there exists a mathematical tool called \textit{Homotopy Theory} designed specifically to tackle such problems. The n'th homotopy group $\pi_n(\mathcal{M})$ classifies qualitatively distinct mappings from the n-th dimensional sphere $\mathcal{S}^n$ into the manifold $\mathcal{M}$.
	
\subsection{The Fundamental Group}
We follow the explanations given by Nakahara\cite{Nakahara}. On a manifold $\mathcal{M}$ we consider closed paths that pass through a point $x$. These paths can be defined by continuous functions $f$ from the interval $0\leq t\leq1$ into $\mathcal{M}$ with the requirement $f(0)=f(1)=x$. Two such closed paths $f$ and $g$ are said to be \textit{homotopic} at $x$ if we can continuously deform one into the other while keeping contact with $x$. 
The space of such closed loops can be further equipped with a product law defined by
\be\label{loopproduct}
\left(f\circ g\right)(t)=
\begin{cases}
f(2t), &0\leq t\leq \frac{1}{2}\\
g(2t-1),&\frac{1}{2}\leq t\leq 1.\\
\end{cases}
\ee

This corresponds to joining the end of the path $f$ to the beginning of the path $g$ thus creating a combined path. This combined path also begins and ends in $x$. The inverse $f^{-1}(t)=f(1-t)$ is recognized merely as walking on $f$ in the opposite direction. We partition the loops by putting all loops homotopic to $g$ in the \textit{homotopy class} $[g]$. Under class multiplication
\be
[f][g]=[f\circ g],
\ee 
which is generalized from the loop product (\ref{loopproduct}), the homotopy classes define a group.
The identity element $[I]$ consists of all loops that are contractible to the point $x$, and the inverse is defined by $[f]^{-1}=[f^{-1}]$.

This group $\pi_1(\mathcal{M},x)$ is called the \textit{fundamental group} of $\mathcal{M}$ at $x$. Definition with respect to a base point $x$ is usually omitted because in a connected space we can define an isomorphism between fundamental groups at any two points by adding a path between the points and its inverse. It then becomes clear that all based groups  $\pi_1(\mathcal{M},x)$ are identical, this allows us to talk about $\pi_1(\mathcal{M})$ - the fundamental group to the manifold $\mathcal{M}$ without the need for a base point.

The actual computation of the fundamental group is greatly simplified if the vacuum manifold $\mathcal{M}$ can be expressed in terms of compact Lie groups. Luckily, this is exactly what happens when some compact Lie group $G$ is spontaneously broken to a smaller subgroup $H$. If we have $\phi_0$, a vacuum expectation value in $\mathcal{M}$ we can generate the remainder of $\mathcal{M}$ by transformations of the form $\phi=D(g)\phi_0$ with $g\in G$ and $g\not\in H$ since $\phi_0$ is invariant to transformations induced by $H$ . Consequently, we can identify any $\phi$ in $\mathcal{M}$ with the coset $gH$. This means the the entire vacuum manifold $\mathcal{M}$ can be identified with the space of cosets of $H$ in $G$, or $\mathcal{M}=G/H$.

We now state, without proof, the \textbf{First Fundamental Theorem} : 
Let $G$ be a connected and simply-connected Lie group, having a subgroup $H$ with a component $H_0$ connected to the identity $e$. We define the quotient group $\pi_0(H)\equiv H/H_0$, $\pi_0(H)$ labels the disconnected components of $H$. The group $\pi_0(H)$ is isomorphic to the fundamental group of the coset space $\pi_1(G/H)$, that is,
\be
\pi_1(G/H)\cong\pi_0(H).
\ee
\subsection{The Second Homotopy Group}
In much the same way as we defined the fundamental group we can define the second homotopy group $\pi_2(\mathcal{M},x)$ as the set of homotopically equivalent classes of maps from the two sphere $\mathcal{S}^2$ into the manifold $\mathcal{M}$. Analogously to the fundamental group, group structure is imposed by considering continuous deformations of two-surfaces that keep the base point $x$ fixed. Two-surfaces that can be continuously shrunk to a point are homotopic to the trivial constant map $I$. In the same fashion as before we can then consider $\pi_2(\mathcal{M})$.

The determination of the second homotopy group is again greatly simplified if the vacuum manifold can be described in terms of a coset space of continuous Lie groups, $\mathcal{M}=G/H$. Given that we take a connected and simply-connected covering group $G$ then the \textbf{\textit{Second Fundamental Theorem}} states that 
\be
\pi_2(G/H)\cong\pi_1(H_0)
\ee
where $H_0$ is a component of the unbroken subgroup $H$ connected to the identity $e$. This reduces the computation of $\pi_2(\mathcal{M})$ to that of $\pi_1(H_0)$.

\clearpage
\section{Introduction to Magnetic Monopoles}

The Goldstone Model (\ref{eq:goldstonelagrangian}) admits a variety of topological defects of different dimensionalities as solutions. We shall discuss a class where $\phi^a$ represents a multicomponent scalar field transforming under a chosen global symmetry group $G$. Monopoles, or point-like defects, arise if the vacuum manifold $\mathcal{M}$ contains a non-contractible two-surface. In other words, the unbroken group $H$ must have a non-trivial fundamental group. This means that any symmetry group $G$ which breaks to leave a $U(1)$ symmetry intact,
\be
G\xrightarrow{SSB}K\times U(1),
\ee
must have monopole solutions since $\pi_1\left(U(1)\right)=\mathbb{Z}$. This means any grand unified theory which breaks to leave a desired residue $U(1)_{EM}$ of electromagnetism must produce monopole solutions.

The simplest example is the 't-Hooft-Polyakov solution\cite{Hooft1974}\cite{Hooft1976}\cite{Polyakov1974}\cite{Polyakov1975} which appears when $SU(2)$ spontaneously breaks to leave $U(1)$ unbroken. Far from the center of the configuration, or core, the three component Higgs field $\phi^a$ takes the form $\phi^a=\eta \frac{x^a}{r}$ i.e. it points radially outwards and has amplitude $\eta$. This configuration is also known as a \textbf{\textit{Hedgehog}}. The gauge fields will align themselves in such a way as to minimize the variational energy. However, a radial magnetic field will remain $\vec{B}=\frac{\vec{r}}{er^3}$ with an overall magnetic flux of $\Phi_B=\frac{4\pi}{e}$.

	Another possibility to consider is that of a global monopole consisting only of the field  $\phi^a=\eta \frac{x^a}{r}$ without the presence of a compensating gauge field. However, the total energy of such objects is linearly divergent. These are consequently strongly confined since the force between monopoles and anti-monopoles(monopoles of opposite charge) is independent of distance.
	
Monopoles which have formed as a result of gauge symmetry breaking carry a unit magnetic flux. This translates to a long range radial magnetic field corresponding to the unbroken symmetry generators. This means that topological monopoles so described are indeed Magnetic Monopoles whose possible existence has been suggested by Dirac\cite{Dirac1931} back in 1931, long before anyone thought to consider spontaneously broken gauge theories.
	Some very good and thorough review papers have been written on the subject of magnetic monopoles, most notable are the ones by Preskill\cite{Preskill1984} and the one by Coleman \cite{Coleman1983}.

\subsection{The Charge Quantization Condition}
\begin{figure}[h!]
\begin{center}
\includegraphics[width=3.3in]{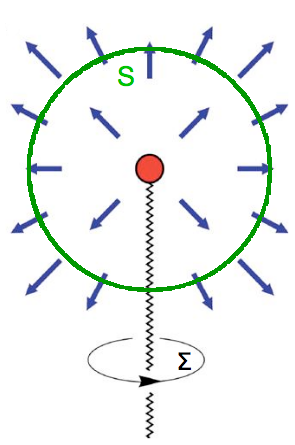}
\caption{\label{Fig.dirac} Illustration of the Dirac string configuration. $\Sigma$ is the surface of a ring that encloses the string. $S$ is the surface of the sphere enclosing the origin, which effectively produces a field of a magnetic monopole. The magnetic field runs along the string and scatters radially when it reaches the end of the string.}
\end{center}
\end{figure}

A relation between the electric charge $e$ and the magnetic charge $g$ was first suggested by Dirac\cite{Dirac1931} who realized that the quantum mechanics of a charged particle in the vicinity of a monopole can not be consistently formulated for arbitrary values of the charges. The action for a charged particle moving in a magnetic field is given by
\be
S=S_0+e\int_{\gamma} \vec{A}\cdot \vec{dx}, 
\ee
where $S_0$ is the free particle action, $\vec{A}$ is the vector potential and $\gamma$ is some path on which the particle is moving. The amplitude for a charged particle to go around a closed path $\Gamma$ is proportional to
\be
\mathcal{A}\propto \text{exp}\{-ie\oint_{\Gamma}\vec{A}\cdot \vec{dx} \}=\text{exp}\{-ie\int_{\Sigma}\vec{B}\cdot \vec{ds} \},\quad \text{for}\quad \Gamma=\partial\Sigma
\ee
 or in other words $\Sigma$ is the surface enclosed by the closed curve $\Gamma$.\\
 For $\mathcal{A}$ to be single valued and independent on the choice of $\Sigma$ the integral must satisfy
 \be
 -e\int_{\Sigma}\vec{B}\cdot \vec{ds} =2\pi n.
 \ee
Dirac introduced an infinite solenoid ending at the origin, as illustrated in figure (\ref{Fig.dirac}). The flux of the Dirac string is of opposite sign to that of a field of the form $\vec{B}=\frac{\vec{gr}}{4\pi r^3}$. Integrating, 
 \be
 -e\int_{\Sigma}\vec{B}\cdot \vec{ds}=e\int_{\text{Sphere}}\vec{B}\cdot \vec{ds} =2\pi n,
 \ee
 we are left with the condition,
\be
e\cdot g=2\pi n.
\ee
This means that the smallest allowed magnetic charge is $g=\frac{2\pi}{e}$. For 't Hooft-Polyakov monopole this condition is satisfied with $n=2$.

\subsection{Monopoles and Grand Unification}
The general condition for the existence of monopoles in a model in which a symmetry group $G$ is broken to leave an unbroken residue $H$ is that the vacuum manifold $\mathcal{M}=G/H$ must contain non-contractible two-surfaces i.e. $\pi_2(G/H)\neq I$. In Grand Unified Theories(GUTs), a semi-simple group $G$ is broken in several stages down to $H_{0}=SU(3)\times U(1)$. But on the other hand we know
\be
\pi_2(G/H)\cong\pi_1(H_{0})\Rightarrow\pi_2(G/H)\cong\pi_1(SU(3)\times U(1))\cong \mathbb{Z}.
\ee
The existence of monopoles is therefore an inevitable prediction of any such GUT. This conclusion is quite trivial and does not depend on the specific group $G$ which we begin with, nor does it depend on the intermediate steps taken in the process of spontaneous symmetry breaking.

In the general case monopoles can have magnetic charges corresponding to several unbroken symmetry generators, for example, in realistic GUTs stable monopoles typically carry also color magnetic charges. However, these charges are usually screened at the characteristic QCD level.

\subsection{Effective Action}\label{ss:nls}
To derive a low energy effective action we begin with a general scalar field Lagrangian $\mathcal{L}(\phi_i,\partial_{\mu}\phi^a)$ which we assume is symmetric under some group $G$ and that there is a well defined minima for the theory(described by some potential) that form a vacuum manifold $\mathcal{M}$ with $\pi_2(\mathcal{M})\neq I$. The dimension  $d=\text{dim}(\mathcal{M})$ is equal to the number of broken generators of $G$. At low energies the massive degrees of freedom are not excited and the field $\phi_i$ is forced to stay on $\mathcal{M}$. In this case the field $\phi_i$ can be parametrized by $d$ fields $\varphi^A$ which can be regarded as coordinates on $\mathcal{M}$:
\be
\phi_i=\phi_i(\varphi^A)\quad,\quad A=(1,...,d).
\ee
The fields $\varphi^A$ may be intuitively associated to goldstone bosons resulting from global symmetry breaking. Putting $\phi_i(\varphi^A)$ into the action we can write an effective action for $\varphi^A(x)$ in the form
\be
S_{\text{scalar}}=\frac{1}{2}\int d^4x\sqrt{-g}G_{AB}(\phi)\partial_{\mu}\varphi^{A}\partial^{\mu}\varphi^{A}\quad,
\ee
where 
\be
G_{AB}=\pd{\phi_i}{\varphi^A}\pd{\phi_i}{\varphi^B}
\ee
is the metric of the vacuum manifold $\mathcal{M}$ and $g$ is the determinant of the complete space-time metric. This action is that of a \textbf{Nonlinear $\sigma$-model}. Alternatively, instead of introducing a new set of independent variables $\varphi^A$ we can work with our original variables $\phi_i$ and force a condition that they must stay on the vacuum manifold as a constraint. For example, in the $SO(N)$ symmetry model the constraint should be
\be
\phi_i\phi_i-\eta^2=0.
\ee
We introduce a Lagrange multiplier $\alpha$. The corresponding Lagrangian is thus
\be
\mathcal{L}_{\phi}=\frac{1}{2}\partial_{\mu}\phi_{i}\partial^{\mu}\phi_{i}-\frac{1}{2}\alpha(\phi_i\phi_i-\eta^2).
\ee
Variation of the fields $\phi_i$ gives
\be
\Box \phi_i+\alpha\phi_i=0
\ee
and variation with respect to $\alpha$ recovers the constraint.

\subsection{The 't Hooft-Polyakov Monopole}
The existence of monopole solutions was first demonstrated independently by 't Hooft\cite{Hooft1974}\cite{Hooft1976} and Polyakov\cite{Polyakov1974}\cite{Polyakov1975} in a gauge model possessing $SO(3)$ symmetry and a Higgs field $\phi^a$ in a triplet representation. The potential was taken to be the mexican hat potential (\ref{eq:mexicanhat}) so that the field acquired a non zero VEV at $|\phi_0|=\eta$. The SO(3) symmetry is spontaneously broken and a U(1) residue remains. The vacuum manifold $\mathcal{M}$ is isomorphic to the two-sphere $S^2$ as $\phi^a$ sits on a sphere of radius $\eta$. The unbroken subgroup $U(1)$ is the group of rotations about $\phi^a$, it can be identified with $U(1)_{EM}$.

The monopole is the simplest topologically non-trivial configuration with a radially pointing Higgs field
\be\label{eq:fieldansatz}
\phi^a=\eta h(r)\frac{x^a}{r}, 
\ee
and the gauge field is taken to be
\be\label{eq:gaugeansatz}
A_i^a=-(1-K(r))\varepsilon^{aij}\frac{x^j}{er^2},\qquad A^{a}_0=0.
\ee
At spatial infinity $\phi$ approaches its VEV and the covariant derivative of $\phi$ must vanish
\be
D_\mu\phi^a=\left(\partial_{\mu}\phi^a-e\varepsilon^{abc}A_{\mu}^{b}\phi^c\right)\xrightarrow{r\rightarrow\infty}0,
\ee
therefore the asymptotic behavior of the functions $h(r)$ and $K(r)$ is given by
\be\label{eq:asy}
h(r)\xrightarrow{r\rightarrow\infty}1,\quad K(r)\xrightarrow{r\rightarrow\infty}0.
\ee
Later we'll see that in the absence of gauge fields the energy energy of the Higgs field (\ref{eq:fieldansatz}) would be linearly divergent at large distances. 
Regularity demands that at the origin $h(0)=0$ and $K(0)=1$. 

A pure gauge field is a field obtained by a gauge transformation on the null field configuration. This field is not pure gauge, even asymptotically. with $h(r)\approx 1$ the field tensor is given by
\be
F_{ij}^a\approx\frac{x^a}{r}\varepsilon_{ijk}\frac{x^k}{er^3},\quad F^a_{0i}=0.
\ee
Note that $F_{ij}^a$ is proportional to $\phi^a$ indicating that it corresponds to an unbroken symmetry generator. 't Hooft suggested the following definition for the electromagnetic field tensor
\be \label{eq:thooftfieldtensor}
\mathcal{F}_{\mu\nu}=\frac{\phi^a}{|\phi|}F_{\mu\nu}^a+\frac{1}{e|\phi|^3}\varepsilon^{abc}\phi^a(D_\mu\phi^b)(D_\nu\phi^c).
\ee
In the gauge where $\phi$ points in the same direction everywhere $\phi=\delta^{a3}|\phi|$ the term quadratic in $A_\mu^a$ drops and (\ref{eq:thooftfieldtensor}) reduces to an elegant $\mathcal{F}_{\mu\nu}=\partial_\mu A_\nu^3-\partial_\nu A_\mu^3$. In the presence of a monopole the insertion of (\ref{eq:fieldansatz}) and (\ref{eq:gaugeansatz}) combined with the conditions (\ref{eq:asy}) into (\ref{eq:thooftfieldtensor}) gives 
\be
\mathcal{F}_{ij}=\varepsilon_{ijk}\frac{x_k}{er^3},\qquad \mathcal{F}_{0i}=0.
\ee
The magnetic field is then given by
\be
\vec{B}=\frac{g\vec{r}}{4\pi r^3},
\ee
with the magnetic charge
\be
g=\frac{4\pi}{e}.
\ee

\subsection{When is $g_{tt}g_{rr}=-1$?}
Let us take a moment and discuss a set of models which exhibit a special feature that allows for a relatively simple calculation of their gravitational fields. A version of this argument was given by Jacobson\cite{Jacobson2007a}. We take models which obey the two conditions:
\begin{itemize}\label{conditions}
\item{The space-time is spherically symmetric, such that any rotation of the coordinates is unnoticeable to a given observer.}
\item{The t-r subspace of the stress energy tensor is proportional to the metric. In other words, models where $(T_{\mu\nu})\propto g_{\mu\nu}$ in the $t-r$ subspace or equivalently with $T_t^t=T_r^r$ and $T_t^r=T_r^t=0$  . Physically this means that the energy density $\rho$ and the radial pressure $p_r$ are related via $\rho=-p_r$. This condition also holds within itself the assumption of time reflection symmetry.}
\end{itemize}
 In such models we notice that the general spherically symmetric metric:
\be\label{eq:genmetric}
ds^2=A(r)dt^2-B(r)dr^2-r^2(d\theta^2+\sin^2{\theta}d\varphi^2),
\ee
reduces to a friendlier:
\be\label{eq:friendly}
ds^2=A(r)dt^2-A(r)^{-1}dr^2-r^2(d\theta^2+\sin^2{\theta}d\varphi^2),
\ee
where the function $A(r)$ can be simply evaluated using the integral
\be\label{Ar}
A(r)=1-\frac{\kappa}{2r}\int_{0}^{r}T_t^t(r')r'^2dr'.
\ee
Strangely, this feature has been mostly absent from standard textbooks, although many writers did notice and use it\cite{Guendelman1996c}\cite{Dymnikova:2003vt}. In the following we will prove this statement.

%

We consider a metric of the form (\ref{eq:genmetric}). The radial light-like  vectors $l^{\mu}$ are defined by $l^{\mu}l_{\mu}=0$. Plugging in the  metric (\ref{eq:genmetric}) we have
\begin{align}
l^{\mu}l_{\mu}&=g_{\mu\nu}l^{\nu}l^{\mu}=g_{tt}l^tl^t+g_{rr}l^rl^r+g_{\theta\theta}l^\theta l^\theta+g_{\phi\phi}l^\phi l^\phi\nonumber\\
&=A(r)l^tl^t-B(r)l^rl^r-r^2l^\theta l^\theta-r^2\sin^2(\theta)l^\phi l^\phi=0.
\end{align}
We see that the radial light-like vectors can be scaled to have components $l^t=B(r)^{1/2}$, $l^r=\pm A(r)^{1/2}$ and $l^{\theta}=l^{\phi}=0$. Looking at the $t-r$ subspace of the Ricci tensor $R^\mu_{\nu}$ and considering time reflection symmetry we can determine $R^t_{r}=R^r_{t}=0$. We are considering situations where $T^t_t=T^r_r$. We conclude that in the $t-r$ subspace the energy stress tensor $T^{\mu}_{\nu}$ is proportional to the unit matrix $(T^{\mu}_{\nu})_{\text(t-r)}\propto\delta_{\nu}^{\mu}$. We lower the index $\mu$ using $g_{\mu\sigma}$ and rename such that $(T_{\mu\nu})_{\text(t-r)}\propto g_{\mu\nu}$. Next, we multiply both sides by $l^{\mu}l^{\nu}$ to get $(T_{\mu\nu})l^{\mu}l^{\nu}\propto g_{\mu\nu}l^{\mu}l^{\nu}=0$, the suffix has been omitted as the radial light-like vectors $l^i$ for $i=\theta, \phi$ vanish. When Einstein's equations $T_{\mu\nu}=\frac{2}{\kappa}\left(R_{\mu\nu}-\frac{1}{2}g_{\mu\nu}R\right)$ hold we can insert the right hand side instead of $T_{\mu\nu}$ and get 
\be
(R_{\mu\nu}-\frac{1}{2}g_{\mu\nu}R)l^{\mu}l^{\nu}=R_{\mu\nu}l^{\mu}l^{\nu}-\frac{1}{2}Rg_{\mu\nu}l^{\mu}l^{\nu}=R_{\mu\nu}l^{\mu}l^{\nu}=0.
\ee
So the problem reduces to that of determining the conditions for the functions $A(r)$ and $B(r)$ such that the term $R_{\mu\nu}l^{\mu}l^{\nu}$ vanishes. 

For the same reasons as the energy-stress tensor, $R_{\mu\nu}$ is diagonal in this subspace. We point out that spherical symmetry implies that $R_{\mu\nu}$ is diagonal in the $\theta-\phi$ subspace as well . We calculate the term $R_{\mu\nu}l^{\mu}l^{\nu}=B(r)R_{tt}+A(r)R_{rr}$ explicitly.
This gives
\be
R_{tt}=\frac{4 A(r) B(r) A'(r)-r B(r) A'(r)^2-r A(r) A'(r) B'(r)+2 r A(r) B(r) A''(r)}{4 r A(r) B(r)^2},
\ee
\be
R_{rr}=\frac{A(r) \left(4 A(r)+r A'(r)\right) B'(r)+r B(r) \left(A'(r)^2-2 A(r) A''(r)\right)}{4 r A(r)^2 B(r)}, 
\ee
so the sum is equal to
\be
R_{\mu\nu}l^{\mu}l^{\nu}=\frac{B(r) A'(r)+A(r) B'(r)}{r B(r)}=\frac{(B(r) A(r))'}{r B(r)}.
\ee
This vanishes if and only if $A(r)B(r)=\text{const}$. A rescaling of the time gives $A(r)B(r)=1$. Thus we have proven that a static, spherically symmetric metric, satisfying $T^{t}_{t}=T_{r}^{r}$ takes our desired form
\be
ds^2=A(r)dt^2-A(r)^{-1}dr^2-r^2(d\theta^2+\sin^2{\theta}d\varphi^2).
\ee

Next we want to prove that the function $A(r)$ is indeed given by (\ref{Ar}). To ease our calculation we shall define a new function $\lambda(r)=\ln A(r)$ so the metric becomes
\be
ds^2=e^{\lambda}dt^2-e^{-\lambda}dr^2-r^2(d\theta^2+\sin^2{\theta}d\varphi^2).
\ee
The $t-t$ component of the Einstein field equations now reads
\be
\frac{\kappa}{2}T_{t}^{t}=-e^{\lambda}\left(\frac{1}{r^2}+\frac{\lambda'}{r}\right)+\frac{1}{r^2},
\ee
rearranging the terms we are left with
\be
(re^{\lambda(r)})'=1-\frac{\kappa}{2}r^2T_{t}^{t}.
\ee
Integrating this and setting the boundary condition $\lambda|_{r=0}=0$ we get 
\be
e^{\lambda(r)}=A(r)= 1-\frac{\kappa}{2r}\int_{0}^{r}T_t^t(r')r'^2dr'.
\ee
Thus we have thus proven our claim. While this seemingly esoteric group of conditions appears to be very restrictive, it turns out that some very important solutions of the Einstein equations satisfy these conditions. These solutions include of course the vacuum where $T_\mu^{\nu}=0$ and also a cosmological constant where $T_\mu^{\nu}\propto\delta_{\mu}^{\nu}$, so both the Schwarzschild metric and the de-Sitter metric fit our criteria. Furthermore, electromagnetic solutions such as the Maxwell or Born-Infeld energy-stress tensors  containing $F_{\mu\alpha}F^{\alpha\nu}$ terms. When contracting these terms with $l^{\mu}l_{\nu}$ these terms yield $V_{\alpha}V^{\alpha}$ where $V_{\alpha}=l^{\mu}F_{\mu\alpha}$.  For a radial magnetic field $V_{\alpha}\propto l_{\alpha}$ hence $V_{\alpha}V^{\alpha}=0$, for a radial magnetic field $F_{\mu\nu}$ is tangential so $V_{\alpha}=0$, therefore also the Reissner-Nordstr$\ddot{\text{o}}$m case fits our conditions.

\subsection{Gravitational Field of a 't Hooft-Polyakov Monopole}\label{ss:gravfieldthooft}
We wish to calculate the gravitational field created by a 't Hooft-Polyakov monopole.  We adapt the calculations of Cho and Freund \cite{Cho1975} and of Bais and Russell \cite{Bais1975}. Starting from (\ref{eq:nonabhiggs}) in Schwarzschild coordinates we shall derive the energy-stress tensor. Then, we shall solve the Einstein field equations to determine the metric generated. We will also note Birkhoff's theorem\cite{Birkhoff1923}\cite{Jebsen1921} that states that every spherically symmetric solution of the Einstein-Maxwell field equations must be stationary and asymptotically flat, hence we shall expect a priori that our solution be of Reissner-Nordstr$\ddot{\text{o}}$m type.

The Lagrangian of our theory is given by (\ref{eq:nonabhiggs}) with the Higgs field $\varphi_i$ now being a triplet of scalar fields $\phi^a$ with $(a=1,2,3)$. We look for spherically symmetric static solutions for which, in Schwarzschild coordinates, the scalar and vector fields take the form (\ref{eq:fieldansatz}) and (\ref{eq:gaugeansatz}) respectively. The spherically symmetric metric reduces to the general form (\ref{eq:genmetric})
\be
ds^2=A(r)dt^2-B(r)dr^2-r^2(d\theta^2+\sin^2{\theta}d\varphi^2),
\ee
We calculate the stress energy tensor by varying the action with respect to the metric,
\be\label{eq:estformula}
T_{\mu\nu}=2\frac{\delta \mathcal{L}}{\delta g^{\mu\nu}}-g_{\mu\nu}\mathcal{L},
\ee
which in our case gives
\be
T_{\mu\nu}=D_{\mu}\phi^aD_{\nu}\phi^a-\frac{1}{2}g_{\mu\nu}g^{\alpha\beta}D_{\alpha}\phi^aD_{\beta}\phi^a-g^{\alpha\beta}F^a_{\mu\alpha}F^a_{\nu\beta}+\frac{1}{4}g_{\mu\nu}g^{\tau\sigma}g^{\alpha\beta}F^a_{\tau\alpha}F^a_{\sigma\beta}+g_{\mu\nu}\frac{1}{4}\lambda\left(\phi^a\phi^a-\eta^2\right)^2.
\ee
We are interested in the gravitational field far away from the core of the monopole. Inserting the asymptotic behavior (\ref{eq:asy}) for the fields we obtain the following expression for the gauge field tensor
\be
F^a_{\mu\nu}=\frac{g}{r^4}\left(-2\varepsilon_{\mu\nu a}r^2-2\varepsilon_{\mu a b} r_b r_\nu+2\varepsilon_{\nu a b}r_b r_{\mu}+\varepsilon_{\mu\nu b}r_a r_b\right),
\ee
where $r_a=(x,y,z)=(r\sin\theta\cos\phi, r\sin\theta\sin\phi, r\cos\theta)$. As stated before, when the Higgs field takes its VEV the covariant derivative, and also the potential, vanishes. The energy-stress tensor reduces to 
\be\label{eq:redest}
T_{\mu\nu}=-g^{\alpha\beta}F^a_{\mu\alpha}F^a_{\nu\beta}+\frac{1}{4}g_{\mu\nu}g^{\tau\sigma}g^{\alpha\beta}F^a_{\tau\alpha}F^a_{\sigma\beta}.
\ee
From previous arguments we expect this energy-stress tensor to satisfy the conditions (\ref{conditions}), we shall see this explicitly. To evaluate the stress-energy tensor we must transform $F^a_{\mu\nu}$ to spherical coordinates. The Field tensor in spherical coordinates can be calculated by, for instance,
 \be
 F^a_{\theta\varphi}=F^a_{xy}\left(\pd{x}{\theta}\pd{y}{\varphi}-\pd{y}{\theta}\pd{x}{\varphi}\right)+F^a_{xz}\left(\pd{x}{\theta}\pd{z}{\varphi}-\pd{z}{\theta}\pd{x}{\varphi}\right)+F^a_{yz}\left(\pd{y}{\theta}\pd{z}{\varphi}-\pd{z}{\theta}\pd{y}{\varphi}\right),
 \ee
 This gives
 \begin{align}\label{eq:schwartzgauge}
 F^x_{\theta\varphi}&=-g\sin^2\theta\cos\varphi\quad ,\nonumber\\
 F^y_{\theta\varphi}&=-g\sin^2\theta\sin\varphi\quad ,\nonumber\\
 F^z_{\theta\varphi}&=-g\sin\theta\cos\theta\quad .
 \end{align}
 And all others vanish. Putting this into the stress energy tensor (\ref {eq:redest}) and using the metric (\ref{eq:genmetric}) we obtain the following result:
 \begin{align}
&T_{tt}=\quad \frac{g^2}{2r^4}e^{2\phi}\qquad,\qquad T_{rr}=-\frac{g^2}{2r^4}e^{2\Lambda}\quad,\nonumber\\
&T_{\theta\theta}=\quad\frac{g^2}{2r^2}\qquad,\qquad T_{\varphi\varphi}=\quad\frac{g^2}{2r^2}\sin^2\theta.
 \end{align}
 We see explicitly that in the $t-r$ subspace the energy-stress tensor is proportional to the metric. This means that the conditions we have set in (\ref{conditions})  hold for this case. The metric then reduces to the form (\ref{eq:friendly})
 \be
 ds^2=A(r)dt^2-A(r)^{-1}dr^2-r^2(d\theta^2+\sin^2{\theta}d\varphi^2).
 \ee
 To determine the gravitational field, we must evaluate the integral \ref{Ar}
\be\label{classical thooft field}
A(r)= 1-\frac{\kappa}{2r}\int_{0}^{r}T_t^t(r')r'^2dr'=1-\frac{A\kappa}{r}+\frac{g^2\kappa}{4r^2}.
\ee

By looking at orbits of neutral test particles or by Birkhoff's theorem, one can determine the integration constant $A$ to be
\be
A=\frac{M_{\text{c}}}{2}\quad,
\ee
where $M_{\text{c}}$ is the monopole core mass. This mass was calculated numerically by Forg$\acute{\text{a}}$cs et al. \cite{Forgacs:2005vx} for different values of the coupling constants.
The metric for a t' Hooft Polyakov monopole in curved space-time is thus given by
\be\label{thooftmetric}
ds^2=\left(1-\frac{M_{\text{c}}\kappa}{2r}+\frac{g^2\kappa}{4r^2}\right)dt^2-\left(1-\frac{M_{\text{c}}\kappa}{2r}+\frac{g^2\kappa}{4r^2}\right)^{-1}dr^2-r^2(d\theta^2+\sin^2{\theta}d\varphi^2).
\ee
We recall that for the physically interesting case the metric takes this form only at large distances. As expected, the geometry is precisely of the Reissner-Nordstr$\ddot{\text{o}}$m form.

\subsection{The Global Monopole}
Monopoles can be formed as a result of a local, as well as a global symmetry breaking. In the latter case we refer to them by the name \textit{Global Monopoles}. Topologically, the conditions for the creation of monopoles are the same in both cases, but their physical properties are quite different. The simplest model that gives rise to a global monopole is described by a Lagrangian similar to that of a Goldstone model (\ref{eq:goldstonelagrangian}) now given by
\be\label{global lagrangian}
\mathcal{L}=\frac{1}{2}\partial_{\mu}\phi^a\partial^{\mu}\phi^a-\frac{1}{4}\lambda(\phi^a\phi^a-\eta^2)^2\quad,
\ee
with $\phi^a$ a scalar field triplet. The model has global $SO(3)$ symmetry which is spontaneously broken into $SO(2)$. This is just a global version of the model discussed before. The simplest solution of this model is the spherically symmetric configuration
\be
\phi^{a}=\eta h(r)\frac{x^a}{r}.
\ee
with 
\be
h(r)\xrightarrow{r\rightarrow\infty}1,\quad h(r)\xrightarrow{r\rightarrow 0}0.
\ee

Outside the core of the monopole where $h(r)\approx1$ the dynamics of the system can be described by a nonlinear $\sigma$-model as described in subsection (\ref{ss:nls}). In this region, the stress energy tensor (\ref{eq:estformula}) can be simply calculated to give
\be\label{eq:estreduced}
T_t^t=T_r^r=\frac{\eta^2}{r^2}\quad,\quad T_{\theta}^{\theta}=T_{\varphi}^{\varphi}=0.
\ee
The total energy or mass is given by
\be\label{eq:monmass}
M_m=4\pi\int_0^R T_{t}^{t}r^2dr=4\pi\eta^2R,
\ee
with $R$ being a cutoff radius. This is linearly divergent as $R$ goes to infinity. This divergence, however, is of little concern as the integral should be cut off at roughly the distance of the nearest anti-monopole.

\subsection{Gravitational Field of a Global Monopole}
The gravitational field produced by a global monopole was first calculated by Barriola and Vilenkin\cite{Barriola1989}.We will recreate the procedure we used to determine the gravitational field of the 't Hooft-Polyakov monopole (\ref{ss:gravfieldthooft}) to determine it ourselves. Starting again with the metric (\ref{eq:genmetric})
\be
ds^2=A(r)dt^2-B(r)dr^2-r^2(d\theta^2+\sin^2{\theta}d\varphi^2),
\ee

The energy stress tensor of the monopole can be calculated from the Lagrangian density (\ref{global lagrangian}) using (\ref{eq:estformula}) and is given generally by
\begin{align}
&T^t_t=\frac{\eta^2h'^2}{2A(r)}+\frac{\eta^2h^2}{r^2}+\frac{1}{4}\lambda\eta^2h(h^2-1)^2,\nonumber\\
&T^r_r=-\frac{\eta^2h'^2}{2A(r)}+\frac{\eta^2h^2}{r^2}+\frac{1}{4}\lambda\eta^2h(h^2-1)^2,\nonumber\\
&T^\theta_\theta=T^{\varphi}_{\varphi}=\frac{\eta^2h'^2}{2A(r)}+\frac{1}{4}\lambda\eta^2h(h^2-1)^2.
\end{align}
The field equations for $\phi^a$ reduce to a single equation for $h(r)$
\be
\frac{1}{A(r)}h''+\left[\frac{2}{A(r)r}+\frac{1}{2B(r)}\left(\frac{B(r)}{A(r)}\right)'\right]h'+\frac{2h}{r^2}-\lambda\eta^2h(h^2-1)=0.
\ee

The problem substantially simplifies outside the core where we can use a nonlinear $\sigma$-model approximation and set $h=1$. In this approximation the energy stress tensor is equal to (\ref{eq:estreduced}). Furthermore, the model satisfies the conditions (\ref{conditions}) therefore its metric reduces to (\ref{eq:friendly}) and we can calculate its gravitational field simply using the general relation (\ref{Ar}) which gives now 
\be\label{classical global field}
A(r)=1-\frac{\kappa}{2}\eta^2-\frac{M_{\text{c}}\kappa}{2r}.
\ee
The mass parameter $M_{\text{c}}$ appears as a constant of integration. To determine its value one has to analyze the behavior of the scalar field inside the core, an order of magnitude estimate gives $M_{\text{c}}\sim\lambda^{-1/2}\eta$. Harari and Lousto \cite{Harari1990d} calculated its value numerically and surprisingly found it to be negative:
\be
M_{\text{c}}\approx-20\lambda^{-1/2}\eta\quad \text{for}\quad 0\leq\frac{\kappa}{2}\eta^2<1.
\ee
For reasonable values of $\lambda$ and $\eta$ - $M_{\text{c}}$ is negligible on any astrophysical scale. By neglecting the mass term we find that the metric in the surroundings of a global monopole takes the form
\be\label{globalmetric}
ds^2=\left(1-\frac{\kappa}{2}\eta^2\right)dt^2-\left(1-\frac{\kappa}{2}\eta^2\right)^{-1}dr^2-r^2(d\theta^2+\sin^2{\theta}d\varphi^2).
\ee
If the monopole satisfies $1-\frac{\kappa}{2}\eta^2>0$ we can rescale $r$ and $t$ and write the monopole metric as
\be
ds^2=dt^2-dr^2-\left(1-\frac{\kappa}{2}\eta^2\right)r^2(d\theta^2+\sin^2{\theta}d\varphi^2).
\ee
This metric describes a space time with a deficit solid angel where the area of a sphere of radius $r$ is not $4\pi r^2$ but is actually smaller, $4\pi\left(1-\frac{\kappa}{2}\eta^2\right)r^2$. 
If the monopole strength is such that $1-\frac{\kappa}{2}\eta^2<0$ a similar rescaling will give
\be
ds^2=-dt^2+dr^2-\left(\frac{\kappa}{2}\eta^2-1\right)r^2(d\theta^2+\sin^2{\theta}d\varphi^2).
\ee
We see that a large monopole strength gives rise to a signature change in the metric. 

Another feature of this metric is that aside from the tiny gravitational effect of the core the monopole exerts no gravitational force on the matter around it. This can be seen by looking at the newtonian potential $\Phi=\frac{\kappa M_c}{2r}=\text{const.}$ since $M_c(r)\sim r$ (\ref{eq:monmass}).
\clearpage
\section{Vacuum Bubble Dynamics}
The dynamics of a generic shell of matter-energy, $\Sigma$, can be described by Israel junction conditions\cite{Israel1966}. This approach was first developed in the pivotal work of Berezin et al.\cite{Berezin:1987bc}.  The shell is the common part of the boundaries of two space-time manifolds $\mathcal{M}_{\pm}$, i.e. $\Sigma\equiv\partial\mathcal{M}_{-}\bigcap\partial\mathcal{M}_{+}$. Geometrically, the embedding of $\Sigma$ in $\mathcal{M}_{\pm}$ is described by the corresponding extrinsic curvatures \footnote{In the following Greek indices $\alpha, \beta, \mu, \nu, ...$ take the values $0, 1, 2, 3$ and Latin indices $a, b, i, j, ...$ take the values $0, 2, 3$.}$K_{ij}^{\pm}$ and, for a non-light-like junction, the junction conditions can be expressed in terms of a jump in said extrinsic curvatures across the shell
\be
[K_{ij}]\equiv K_{ij}^{(+)}-K_{ij}^{(-)},
\ee 
which is related to the shell's energy-stress tensor $S_{ij}$\cite{Blau1987}\cite{Ansoldi2007} by 
\be
[K_{ij}]=\frac{\kappa}{2}S_{ij}.
\ee
Furthermore, the conservation equations imply
\be
S^{i}_{j;i}=[e^{\alpha}_{(j)}T^{\beta}_{\alpha}n_{\beta}],
\ee
where $T_{\mu\nu}^{(\pm)}$ describes the energy-stress tensors of $\mathcal{M}_{\pm}$, $n^{\mu}$ is the normal to the shell $\Sigma$, which we assume to be pointing from $\mathcal{M}_{-}$ to $\mathcal{M}_{+}$, and $e_{(i)}^{(\pm)\alpha}$ are the components of a basis in the tangent space to $\Sigma$ when evaluated in $\mathcal{M}_{\pm}$, respectively.

We assume spherically symmetric solutions of the form (\ref{eq:friendly}) and study the dynamics of a true vacuum bubble circumscribed by a false vacuum region. Between the two regions we assume a thin mass shell which has a proper surface tension. The metric ansatz for the inside and the outside regions, respectively, are
\be
ds^2_\text{i}=A_\text{i}(r)dt^2-A_\text{i}(r)^{-1}dr^2-r^2(d\theta^2+\sin^2{\theta}d\varphi^2),
\ee
and
\be
ds^2_\text{o}=A_\text{o}(r)dt^2-A_\text{o}(r)^{-1}dr^2-r^2(d\theta^2+\sin^2{\theta}d\varphi^2).
\ee
We shall take the inside vacuum region to have a zero energy density, this makes the interior newtonian potential $A_\text{i}(r)$ equal to $1$. The exterior metric would change depending on the model we are concerning.

The metric ansatz for transition region is
\be
ds^2_{\text{shell}}=d\tau^2-R^2(\tau)d\Omega^2,
\ee
where $R(\tau)=r(\pm)|_{\Sigma}$. The continuity of the three metric across $\Sigma$ is then realized, as requested by the Israel junction conditions. Within the above settings the only remaining junction condition, mentioned above, can be written as
\be\label{hgvhg}
\epsilon_{(-)}\sqrt{\dot{R}^2+A_\text{i}(R)}-\epsilon_{(+)}\sqrt{\dot{R}^2+A_\text{o}(R)}=\frac{\kappa M(R)}{16\pi R}, 
\ee
where $M(R)$ is the mass of the shell. We take the shell to be comprised of non-exotic matter with a positive constant surface tension $\sigma$. This gives us a total mass of $M(R)=4\pi R^2\sigma$. The quantities $\epsilon_{\pm}$ are signs, related to the direction of the normal in the maximal extension of the space-times described by the metrics $g_{\mu\nu}^{\pm}$. The equation
\be\label{jump}
\epsilon_{(-)}\sqrt{\dot{R}^2+A_\text{i}(R)}-\epsilon_{(+)}\sqrt{\dot{R}^2+A_\text{o}(R)}=KR, 
\ee
where $K\equiv\frac{\kappa\sigma}{4}$. Squaring this term, rearranging, and squaring again, we can reduce equation (\ref{jump}) to the form
\be
\dot{R}^2+V_{\text{eff}}(R)=0,
\ee
where the \textbf{\textit{effective bubble potential}} is given by
\be
V_{\text{eff}}(r)=A_\text{o}(r)-\frac{(A_\text{i}(r)-A_\text{o}(r)-K^2r^2)^2}{4K^2r^2}.
\ee
This is essentially a classic equation of motion for a particle with $"\text{mass}"=2$ moving inside the potential $V_{\text{eff}}$ which has a total energy $E_{\text{particle}}=0$.

\subsection{Bubble Dynamics in a t' Hooft-Polyakov Surrounding}\label{bubbletmtthooft}
We look at the dynamics of a true vacuum bubble circumscribed by the metric (\ref{thooftmetric}) which as we saw describes the gravitational field created by a 't-Hooft Polyakov monopole. We take the newtonian potentials to be
\be
A_\text{i}(r)=1,
\ee
and
\be
A_\text{o}(r)=1-\frac{2M}{r}+\frac{Q_g^2}{r^2},
\ee
where $Q_g=g^2\kappa/4$ represents the charge of the monopole and $M$ is the effective combined mass of the shell and the monopole in natural units and is a free parameter. These terms give the shell an effective potential given by
\begin{align}
V_{\text{eff}}(r)=&1+\frac{Q_g^2}{r^2}-\frac{2 M}{r}-\frac{\left(-\frac{Q_g^2}{r^2}+\frac{2 M}{r}-K^2 r^2\right)^2}{4 K^2 r^2}=1-\frac{\left(Q_g^2-2 M r-K^2 r^4\right)^2}{4 K^2 r^6}.
\end{align}
This potential is dependent on three variables, $Q_g$ and $M$ as defined before and $K$, which represents the tension of the vacuum bubble shell. The equation $V_{\text{eff}}(r)=0$ is a eighth order equation in $r$. It admits a maximum of eight real roots from which four may be positive. When $r\rightarrow 0$ and $Q_g\neq0$ the function $V_{\text{eff}}(r)\sim -1/r^6\rightarrow -\infty$, when $r\rightarrow \infty$ the potential asymptotically behaves like $V_{\text{eff}}(r)\sim -r^2$ in the presence of any non-zero surface tension in the shell. We assume $K$ to be positive.

\begin{figure}[h!]
\begin{center}
\includegraphics[width=3.3in]{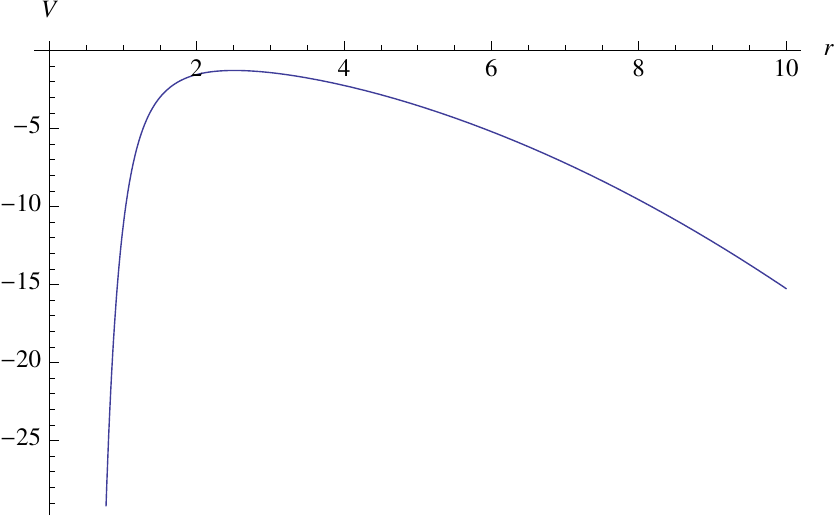}

\caption{\label{Fig:V_eff free} The effective bubble potential describing the \textit{free solution} for the 't Hooft-Polyakov monopole gravitational field surroundings. This is the form of the solution when the monopole charge $Q_g$ is small, the mass of the shell $M$ is large and the surface tension is some positive constant. We see that classical solutions are allowed everywhere and the shell will either expand indefinitely or collapse depending on initial velocity.}
\end{center}
\end{figure}

\begin{figure}[h!]
\begin{center}
\includegraphics[width=3.3in]{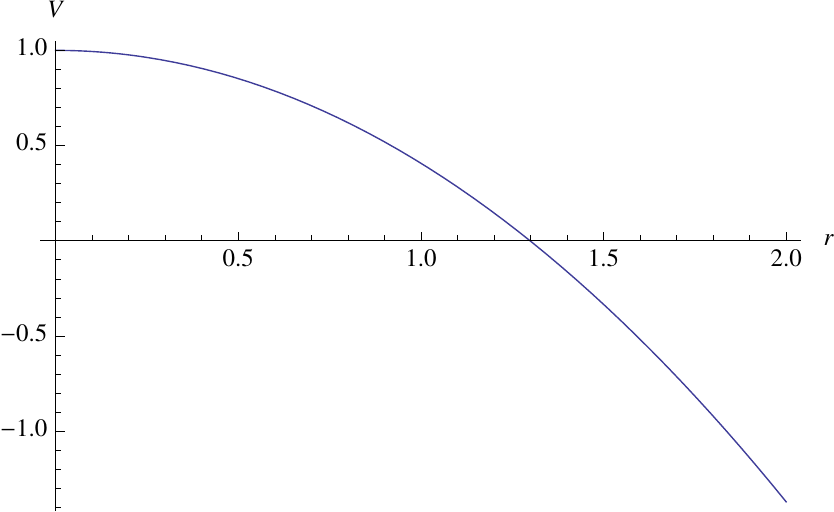}

\caption{\label{Fig:V_eff Q=0} The effective bubble potential describing the Coleman-de Luccia \textit{bounce solution} for the 't Hooft-Polyakov monopole gravitational field surroundings. This is the form of the solution when the monopole charge $Q_g=0$, $M=0$ and the surface tension is some positive constant. We see that classical solutions can be found only above some minimal value of the radius $r_{\text{min}}$. In general if a shell begins with some $r>r_{\text{min}}$ and is shrinking, it will keep shrinking until $r=r_{\text{min}}$, then re-expand indefinitely.}
\end{center}
\end{figure}

\begin{figure}[h!]
\begin{center}
\includegraphics[width=3.3in]{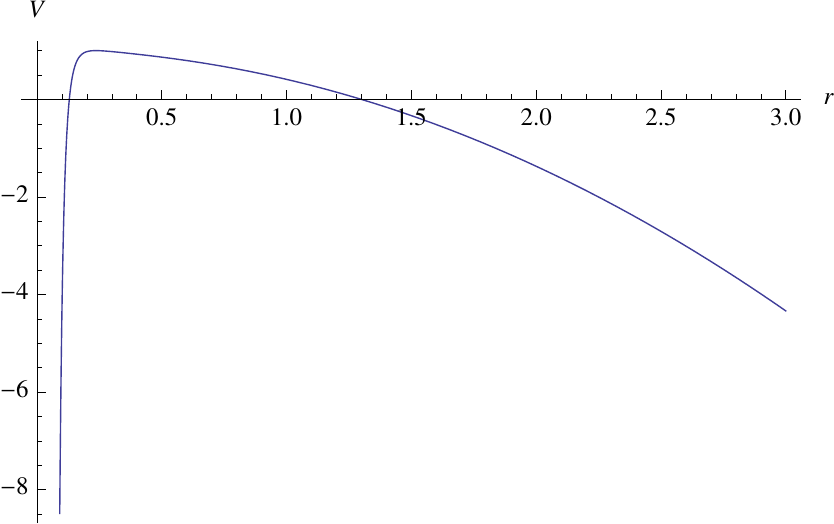}

\caption{\label{Fig:thooftsmallQ} The effective bubble potential in a 't Hooft-Polyakov monopole gravitational surroundings, For for a non-zero monopole charge $Q_g$, zero mass $M$ and some positive surface tension. We see that classically the shell has two allowed regions. For small radii the shell will eventually collapse into itself, for large ones the shell will expand indefinitely. }
\end{center}
\end{figure}

\begin{figure}[h!]
\begin{center}
\includegraphics[width=3.3in]{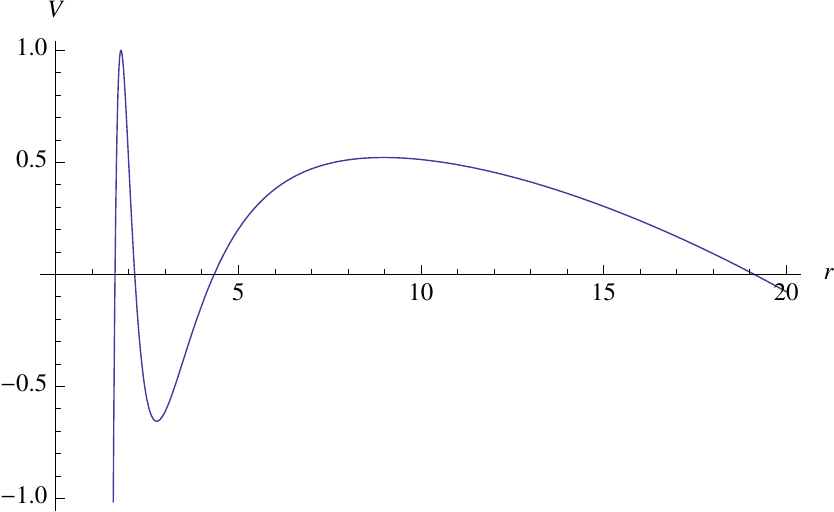}

\caption{\label{Fig:localbreathing} The effective bubble potential in a 't Hooft-Polyakov monopole gravitational surroundings, For for a non-zero monopole charge $Q_g$ and non-zero mass $M$ with some positive surface tension. We see that classically the shell has three allowed regions. For small radii the shell will eventually collapse into itself, for large ones the shell will expand. However for radii in the breathing region the shell will oscillate indefinitely.}
\end{center}
\end{figure}

We therefore have four types of solutions
\begin{itemize}\label{solutionsthooft}
\item{No real positive roots - a \textit{free solution}, where the shell may either grow indefinitely or collapse depending on its initial velocity, although the velocity of the shell itself does change during these processes. This solution is shown in figure (\ref{Fig:V_eff free}).

}
\item{One real positive root - For $Q_g=0, M=0$ we basically have two true vacuum regions separated by a spherical shell. This is the usual Coleman-de Luccia \textit{bounce solution}\cite{Coleman:1980aw} depicted in figure (\ref{Fig:V_eff Q=0}) , were we have a classical solution that contracts, reaches a minimum radius and then re-expands.

}
\item{Two real positive roots - For a small monopole strength $Q_g$ there is a rage of masses $M$ where our system has two classical solutions depicted in figure (\ref{Fig:thooftsmallQ}). 
\begin{enumerate}
\item{A \textit{collapsing solution}, which starts evolving at zero radius and, after reaching a maximum radius of expansion, collapses back to zero radius;}
\item{A \textit{bouncing solution} where as before, the shell starts evolving with an infinite radius, shrinks to some finite minimal radius, then re-expands indefinitely.}
\end{enumerate}
}
\item{Four real positive roots - There exists a range in the parameter space where for a positive mass $M$ there could be four real roots to the equation. This means that along with the collapsing and bouncing solutions described above there is a third possibility. These are solutions where if the shell's initial radius is between some maximal and minimal values i.e. $r_{\text{max}}>r_{\text{initial}}>r_{\text{min}}$
 the shell will classically oscillate between those two values. We call these solutions \textit{breathing bubbles}. We call the region $r_{\text{max}}>r_{\text{initial}}>r_{\text{min}}$ the \textit{breathing region}.  
Classically, these solutions are of little interest because there is little reason to assume an initial condition inside the breathing region. However, if we consider semiclassical effects we see that even a shell beginning at very large or very small radii may transit to the breathing region via a tunneling process. Notice however that this phenomena only happens for a massive shell.This is depicted in figure (\ref{Fig:localbreathing}).
}
\end{itemize}

\clearpage

\subsection{Bubble Dynamics in a Global Monopole Surroundings}\label{bubbletmtglobal}
\begin{figure}[h]
\begin{center}
\includegraphics[width=3.3in]{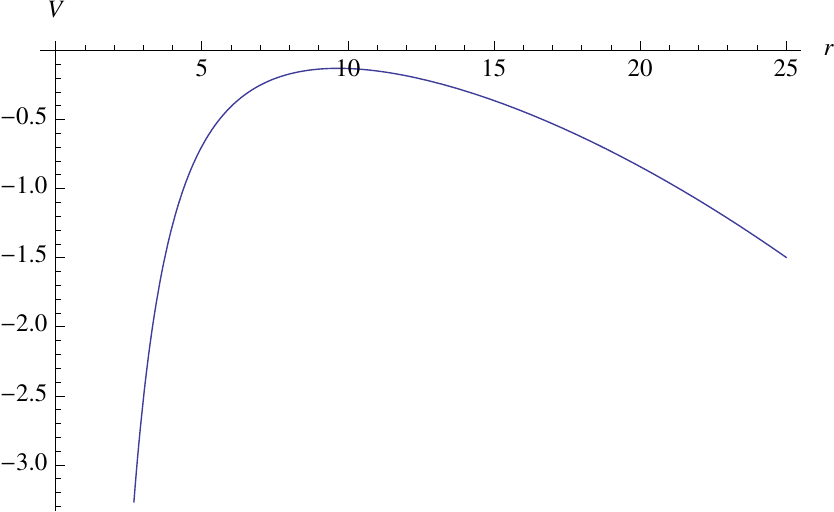}

\caption{\label{Fig:h>1} 
The effective bubble potential in a global monopole gravitational surroundings, For for a monopole strength in the range $h\geq1$ and some positive surface tension. We see that all radii are classically allowed for the bubble shell. depending on initial velocity, the shell will either collapse or expand indefinitely.}
\end{center}
\end{figure}

\begin{figure}[h]
\begin{center}
\includegraphics[width=3.3in]{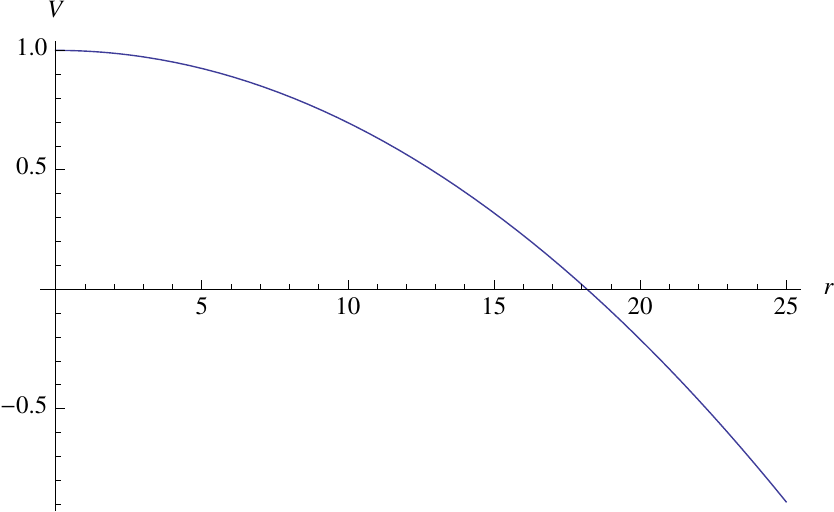}

\caption{\label{Fig: globaleta0} The effective bubble potential describing the Coleman-de Luccia \textit{bounce solution} for the global monopole gravitational field surroundings. This is the form of the solution when the monopole strength $h=0$, its mass $M=0$, and the surface tension is some positive constant. We see that classical solutions can be found only above some minimal value of the radius $r_{\text{min}}$. In general if a shell begins with some $r>r_{\text{min}}$ and is shrinking, it will keep shrinking until $r=r_{\text{min}}$, then re-expand indefinitely.}
\end{center}
\end{figure}

\begin{figure}[h]
\begin{center}
\includegraphics[width=3.3in]{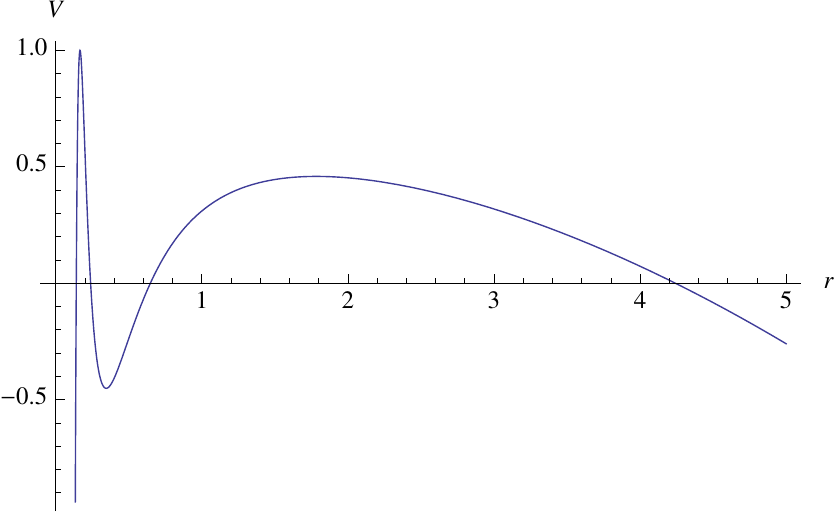}

\caption{\label{Fig:globalfour} 
The effective bubble potential in a global monopole gravitational surroundings, For a monopole strength in the range $0<h<1$ and a \textbf{negative mass} $M$. We see that classically the shell has three allowed regions. For small radii the shell will eventually collapse into itself and for large ones the shell will expand indefinitely, but for radii in the range $r_{max}>r>r_{min}$ the shell will classically oscillate indefinitely.}
\end{center}
\end{figure}

Turning our attention to the global case we look for solutions where vacuum bubble is circumscribed by metric is given by (\ref{globalmetric}). The newtonian potentials read now
\be
A_\text{i}(r)=1,
\ee
and
\be
A_\text{o}(r)=1-h-\frac{2M}{r},
\ee
where $h=\frac{\kappa}{2}\eta^2$ represents the strength of the monopole and $M$ again represents the effective mass of the shell in natural units. These terms give the shell an effective potential given by
\be
V_{\text{eff}}(r)=1-h-\frac{2 M}{r}-\frac{\left(h+\frac{2 M}{r}-K^2 r^2\right)^2}{4 K^2 r^2}.
\ee
The equation $V_{\text{eff}}(r)=0$ is of degree six and may give rise to a maximum of six real roots, of which four may be positive. The asymptotic behavior for $r\rightarrow 0$ is dominated by the $\sim -1/r^4$ term proportional to $M$, hence $V_{\text{eff}}(r)\xrightarrow{r\rightarrow0}-\infty$. On the other limit when $r\rightarrow\infty$ the behavior of the function is dominated by the $\sim -r^2$ term proportional to $K$. This means that the potential again admits four types of solutions  (\ref{solutionsthooft}). We assume $K$ to be positive.
\begin{itemize}
\item{No real positive solution - For $h\geq 1$  and any positive mass $M$ we get a \textit{free solution}, where the shell may either grow indefinitely or collapse depending on its initial velocity, although the velocity of the shell itself does change during these processes. This solution is shown in figure (\ref{Fig:h>1}).}
\item{One real positive solution - For zero monopole strength $h$ and mass $M$ , there is no monopole and we again get the Coleman-de Luccia \textit{bounce solution} depicted in figure (\ref{Fig: globaleta0}) as described above.}
\item{Two real positive solutions - For $1>h>0$ and any positive mass $M$ we get both a \textit{collapsing solution} if the initial radius is small and a \textit{bounce solution} if the initial radius is large.}
\item{Four real positive solutions - For $1>h>0$ there is a range of \textbf{negative masses} $M$ in which we get four positive real solutions to the equation. This again translates to the appearance of a \textit{breathing region}. This is depicted in figure (\ref{Fig:globalfour}).}
\end{itemize}

\begin{figure}[h]
\begin{center}
\includegraphics[width=3.3in]{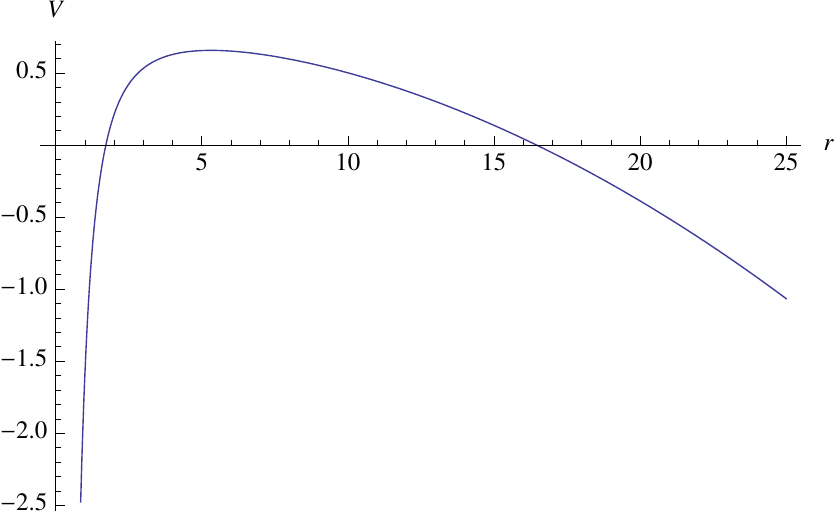}

\caption{\label{Fig:h<1} 
The effective bubble potential in a global monopole gravitational surroundings, For a monopole strength in the range $0<h<1$, with a positive mass $M$ and some positive surface tension. We see that classically the shell has two allowed regions. For small radii the shell will eventually collapse into itself, for large ones the shell will expand indefinitely.}
\end{center}
\end{figure}

\clearpage
\section{Introduction to the Two Measure Field Theory}
\subsection{Motivation to a Two Measure Field Theory: The Volume Measure of the Space-time Manifold and its Dynamical Degrees of Freedom}
\label{sec:Motivation}
One of the motivations for using an additional measure of integration in the action principle is closely related to a possible degeneracy of the metric\cite{Guendelman2008k}. Solutions with degenerate metric were a subject of long-standing discussions starting probably with the paper by Einstein and Rosen\cite{Einstein1935a}. In spite of the difficulty of interpreting the physical meaning of such solutions in the classical theory of gravitation, the prevailing view is that these solutions due have physical meanings and therefore must be included in the path integral\cite{Hawking:1979zw}\cite{Tseytlin:1981ks}\cite{D'Auria:1981yg}. If we then appropriately extend the notions of general relativity using a first order formalism, solutions with $g(x)=\text{det}(g_{\mu\nu})=0$ allow to describe changes in the space-time topology\cite{Hawking:1979zw}\cite{Horowitz:1990qb}. There are known also classical solutions with a change of the signature of the metric tensor. Using terminology by Tseytlin\cite{Tseytlin:1981ks}, the space-time regions with $g(x)=0$ can be treated as having \textit{metrical dimension} $D<4$.

The simplest example of a degenerate metric is $g_{\mu\nu}=0$ and has an arbitrary affine connection. Such solutions have been studied by D'Auria and Regge\cite{D'Auria:1981yg}, Tseytlin\cite{Tseytlin:1981ks},  Horowitz\cite{Horowitz:1990qb} and others. It has been suggested that $g_{\mu\nu}=0$ should be interpreted as basically a non-classical phase in which diffeomorphism invariance is unbroken, this was suggested to be realized at high temperature and curvature.

The question to ask is this: Does the condition $g(x)=0$ necessarily mean that in the neighborhood of $x$ the dimension of the space-time manifold becomes $D<4$? At a first sight it seems to be so because we define the volume element to be
\be
dV_{\text{metrical}}=\sqrt{-g}d^4x.
\ee
But if we were to look back at the procedure resulting in this volume element we are to realize that this definition was possible only after our four dimensional manifold $\mathcal{M}_4$ was equipped with a metric structure. For this reason the suffix 'metrical' was added here. However, even with no defined metric, we can still define a non-zero volume element to the manifold $\mathcal{M}_4$ in 4 dimensions. The accepted way of doing that consists of the construction of a differential 4-form using for example four differential 1-forms that we shall call $d\varphi_a\quad(a=1,2,3,4)$, this results in the 4-form $d\varphi_1\wedge d\varphi_2\wedge d\varphi_3\wedge d\varphi_4$ . Each of these 1-forms may be defined by some scalar field $\varphi_a$ residing on the manifold. The appropriate volume element of the four dimensional manifold $\mathcal{M}_4$ can be represented in the following way:
\be\label{eq:fourform}
dV_{\text{manifold}}=4!d\varphi_1\wedge d\varphi_2\wedge d\varphi_3\wedge d\varphi_4=\Phi d^4x,
\ee
where
\be\label{eq:manmeasure}
\Phi=\varepsilon_{abcd}\varepsilon^{\mu\nu\lambda\sigma}(\partial_{\mu}\varphi_a)(\partial_{\nu}\varphi_b)(\partial_{\lambda}\varphi_c)(\partial_{\sigma}\varphi_d)
\ee
is a volume measure which unlike the standard measure, $\sqrt{-g}$, is independent of the metric $g_{\mu\nu}$. This is emphasized by the suffix 'manifold', we will call $\Phi$ - a \textbf{\textit{manifold volume measure}}.

If $\Phi(x)$ is non-zero, one can think of the four scalar fields $\varphi_a$ as describing a homeomorphism from an open neighborhood of the point $x$ to the four dimensional Euclidean space $\mathcal{R}^4$. However if one allows dynamical mechanisms to degenerate the metric and hence reduce the metrical dimension, there is no apparent reason to ignore a similar effect on the manifold measure $\Phi$. This possibility of a vanishing, or even worse, a sign-changing manifold volume measure seems to be here more natural, since the manifold volume measure $\Phi$ is sign indefinite. We also note that there is generically no obligation for the metric and manifold volume measures to vanish simultaneously.

The idea to use differential forms to describe dynamical degrees of freedom of the space-time differential manifold first originated in Taylor's attempt to quantize gravity\cite{Taylor:1978fy}. Taylor argued that quantum mechanics is not compatible with the Riemannian metric space-time. Moreover, in the quantum regime space-time is not even an affine manifold. Only in the classical limit the metric and connections emerge, and only then one is able to construct a traditional space-time description. In his paper \cite{Taylor:1978fy} Taylor describes the transition to the classical limit rather in the form of a general prescription. We would like to pay attention to a possibility that was ignored in \cite{Taylor:1978fy}. In principle it is possible that in the classical limit not only have the metric and connection emerged, but also some of the differential forms have kept(or restored) certain dynamical effects. In such a case, the traditional description of space-time may occur to be incomplete. The key idea of the proposed theory is that one of these "lost" differential forms, the 4-form (\ref{eq:fourform}), survives in the classical limit as describing dynamical degrees of freedom of the volume measure of the space-time manifold. and hence can affect the gravity theory in the classical level too.

The addition of four scalar fields $\varphi_a(x)$ as new variables to the usual set of variables(degrees of freedom of the metric, connection, matter etc.) which also abide action principles leads us to expect an effect of gravity and matter on the manifold volume measure $\Phi$ and vice versa.

Another interesting point is the issue of the orientability of the manifold $\mathcal M_4$. It is known that for a manifold of dimension $n$ to be orientable it must posses a \textbf{\textit{volume form}} - a differential form of degree $n$ which is nonzero at every point on the manifold. Therefore two possible signs of the manifold volume measure (\ref{eq:manmeasure}) are associated with two possible orientations of the space-time manifold. This means that besides the dimensional reduction and topology changes on the level of the differential manifold, the incorporation of the manifold volume measure $\Phi$ allows to realize solutions describing a dynamical change of the orientation of the space-time manifold.

The simplest way to realize the existence of a second volume measure is to modify the action which should now consist of two terms, one with the usual measure $\sqrt{-g}$ and another with the measure $\Phi$,
\be\label{eq:TMTaction}
\mathcal{S}=\int\left(\Phi L_1+\sqrt{-g}L_2\right)d^4x,
\ee
where two Lagrangians $L_1$ and $L_2$ coupled to manifold and metrical volume elements respectively appear. This is the action of the Two Measure Field Theory(TMT). Our experience with TMT models shows us that at least at a classical level, the Lagrangians $L_1$ and $L_2$ are independent of the scalar fields $\varphi_a(x)$ so we proceed with this basic assumption. This means that manifold volume measure degrees of freedom enter into TMT only through the manifold volume measure $\Phi$. In such a case, the action (\ref{eq:TMTaction}) possesses an infinite dimensional symmetry
\be
\varphi_a\rightarrow\varphi_a+f_a(L_1),
\ee
where $f_a(L_1)$ are arbitrary functions of $L_1$. This gives an encouraging clue for the absence of scalar field $\varphi_a(x)$ in $L_1$ and $L_2$ after quantum effects are taken into account.

Notice that (\ref{eq:TMTaction}) is just a convenient ordering of the action terms in a general form. In concrete models, the action (\ref{eq:TMTaction})  can always be rewritten in an equivalent form such that each term in the action has its own single volume measure which is some linear combinations of the two measures $\sqrt{-g}$ and $\Phi$.

\subsection{Classical Equations of Motion}
We next try to come up with a general prescription for the equations of motion in the theory. Working in first order formalism we begin our calculations by varying the measure fields $\varphi_a$, we get $B_a^{\mu}\partial_\mu L_1=0$ where $B^{\mu}_{a}=\varepsilon^{\mu\nu\alpha\beta}\varepsilon_{abcd}
\partial_{\nu}\varphi_{b}\partial_{\alpha}\varphi_{c}
\partial_{\beta}\varphi_{d}.$
Since $\det (B^{\mu}_{a}) = \frac{4^{-4}}{4!}\Phi^{3}$ it follows
that if $\Phi\neq 0$,
\begin{equation}\label{eq:constraint0}
 L_{1}=sM^{4} =\text{const.}
\end{equation}
where $s=\pm 1$ and $M$ is a constant of integration with the
dimension of mass. 

Variation of the action with respect to the metric $g_{\mu\nu}$ yields
\be\label{eq:metricvar}
\zeta\pd{L_1}{g^{\mu\nu}}+\pd{L_2}{g^{\mu\nu}}-\frac{1}{2}g_{\mu\nu}L_2=0,
\ee
where we have defined
\be\label{eq:zeta}
\zeta\equiv\frac{\Phi}{\sqrt{-g}},
\ee
to be the scalar field build from the ratio of the scalar densities $\Phi$ and $\sqrt{-g}$.

We study here models with the Lagrangians of the form:
\be\label{eq:classmodel}
L_1=-\frac{1}{\kappa}R(\Gamma,g)+L_1^{\text{m}}\quad,\quad L_2=-\frac{b_g}{\kappa}R(\Gamma,g)+L_2^{\text{m}},
\ee
where $\Gamma$ stands for affine connection, $R(\Gamma,g)=g^{\mu\nu}R_{\mu\nu}(\Gamma)$, $R_{\mu\nu}(\Gamma)=R^\lambda_{\mu\nu\lambda}(\Gamma)$ and the Riemann curvature tensor is given by
\be\label{Riemann}
R^\lambda_{\mu\nu\sigma}(\Gamma)=\Gamma^\lambda_{\mu\nu,\sigma}-\Gamma^\lambda_{\mu\sigma,\nu}+\Gamma^{\lambda}_{\alpha\sigma}\Gamma^{\alpha}_{\mu\nu}-\Gamma^{\lambda}_{\alpha\nu}\Gamma^{\alpha}_{\mu\sigma}.\ee
  The dimensionless factor $b_g$ in front of $R(\Gamma,g)$ in $L_1$ appears because there is no reason to assume that the coupling constants of scalar curvature to the two measures be equal. We choose $b_g>0$ and $\kappa=16\pi G$ where $G$ is newtons constant. $L_1^{\text{m}}$ and $L_2^{\text{m}}$ are the matter Lagrangians that may include all possible terms included in standard field theory models with the sole measure $\sqrt{-g}$.

Since the measure $\Phi$ is sign indefinite, the total volume measure $(\Phi+b_g\sqrt{-g})$ in the gravitational term $-\kappa^{-1}\int R(\Gamma,g)(\Phi+b_g\sqrt{-g})d^4x$ is also generically sign indefinite.

Variation of the action with respect to the connection coefficients $\Gamma_{\mu\nu}^{\lambda}$ ,we get
\be
-\Gamma_{\mu\nu}^{\lambda}-\Gamma_{\beta\mu}^{\alpha}g^{\beta\lambda}g_{\alpha\nu}+\delta_{\nu}^{\lambda}\Gamma_{\mu\alpha}^{\alpha}+\delta^{\lambda}_{\mu}g^{\alpha\beta}\Gamma_{\alpha\beta}^{\gamma}g_{\gamma\nu}-g_{\alpha\nu}\partial_{\mu}g^{\alpha\lambda}+\delta_{\mu}^{\lambda}g_{\alpha\nu}\partial_{\beta}g^{\alpha\beta}-\delta_{\nu}^{\lambda}\frac{\Phi,_{\mu}}{\Phi}+\delta_{\mu}^{\lambda}\frac{\Phi,_{\nu}}{\Phi}=0.
\ee
We will look for solutions of the form
\be
\Gamma_{\mu\nu}^{\lambda}=\{_{\mu\nu}^{\lambda} \}+\Sigma_{\mu\nu}^{\lambda},
\ee
where $\{_{\mu\nu}^{\lambda} \}$ are the Christoffel's connection coefficients. The $\Sigma_{\mu\nu}^{\lambda}$ satisfies the equation
\be\label{eq:Sigma}
-\sigma,_{\lambda}g_{\mu\nu}+\sigma,_{\mu}g_{\nu\lambda}-g_{\nu\alpha}\Sigma^{\alpha}_{\lambda\mu}-g_{\mu\alpha}\Sigma^{\alpha}_{\nu\lambda}+g_{\mu\nu}\Sigma^{\alpha}_{\lambda\alpha}+g_{\nu\lambda}g_{\alpha\mu}g^{\beta\gamma}\Sigma^{\alpha}_{\beta\gamma}=0,
\ee
where $\sigma=\ln{(\zeta+b_g)}$, zeta is the ratio of the measures which as defined before (\ref{eq:zeta}) given by $\zeta\equiv\frac{\Phi}{\sqrt{-g}}$. The general solution of equation (\ref{eq:Sigma}) is
\be
\Sigma_{\mu\nu}^{\alpha}=\delta^{\alpha}_{\mu}\lambda,_{\nu}+\frac{1}{2}\left(\sigma,_{\mu}\delta_{\nu}^{\alpha}-\sigma,_{\beta}g_{\mu\nu}g^{\alpha\beta}\right)
\ee
where $\lambda$ is an arbitrary function which appears due to the existence of the Einstein-Kaufman $\lambda$ symmetry: The curvature tensor (\ref{Riemann}) is invariant under the $\lambda$ transformation
\be
\Gamma_{\mu\nu}^{'\alpha}(\lambda,\sigma)=\Gamma_{\mu\nu}^{\alpha}+\delta_{\mu}^{\alpha}\lambda,_{\nu}.
\ee
This symmetry turns out to be quite useful in our case. If we choose $\lambda=\sigma/2$, then the antisymmetric part of $\Sigma^{\alpha}_{\mu\nu}$ disappears and we get
\be
\Sigma_{\mu\nu}^{\alpha}=\frac{1}{2}\left(\delta_{\mu}^{\alpha}\sigma,_{\nu}+\delta^{\alpha}_{\nu}\sigma,_{\mu}-\sigma,_{\beta}g_{\mu\nu}g^{\alpha\beta}\right)
\ee
which contributes to the nonmetricity. Summing up this calculation yields
\be\label{eq:connection}
\Gamma_{\mu\nu}^{\lambda}=\{_{\mu\nu}^{\lambda} \}+\frac{1}{2}\left(\delta_{\mu}^{\alpha}\sigma,_{\nu}+\delta^{\alpha}_{\nu}\sigma,_{\mu}-\sigma,_{\beta}g_{\mu\nu}g^{\alpha\beta}\right),
\ee

We notice that $\sigma,_{\mu}=\zeta,_{\mu}/(\zeta+b_g)$ so that if $\zeta\neq\text{const.}$ the covariant derivative of $g_{\mu\nu}$ with this connection is non-zero (nonmetricity) and consequently the geometry of the space-time with metric $g_{\mu\nu}$ is generically non-Riemannian. The gravity and matter field equations obtained by means of the first order formalism contain both $\zeta$ and its gradient. It turns out that at least on a classical level, the measure fields $\phi_a$ affect the theory only through the scalar field $\zeta$.

For the class of models we are considering (\ref{eq:classmodel}), the consistency of the constraint (\ref{eq:constraint0}) and the gravitational equations (\ref{eq:metricvar}) has the form of the following constraint
\be\label{eq:constraint1}
(\zeta-b_g)(sM^4-L_1^{\text{m}})+g_{\mu\nu}\left(\zeta\pd{L_1^{\text{m}}}{g^{\mu\nu}}+\pd{L_2^{\text{m}}}{g^{\mu\nu}}\right)-2L_2^{\text{m}}=0,
\ee
which determines $\zeta(x)$(up to the chosen value of the integration constant $sM^4$) as a local function of matter fields and metric. Note that does not abide by any dynamical equation of motion and its space-time behavior is totally determined by the matter fields and metric via the constraint (\ref{eq:constraint1}). With this in mind, and by observing that $\zeta$ enters into all equations of motion, it generically will have straightforward effects on dynamics of the matter and gravity through the form of potentials, variable masses and self interactions \cite{Guendelman:1998ms,Guendelman:2004jm,Guendelman:2004bn,Guendelman2006c,Guendelman2007,Guendelman2008k}.

To understand the structure of TMT, it is important to note that TMT(where, as we suppose, the scalar fields $\varphi_a$ enter only via the measure $\Phi$) is a constrained dynamical system. In fact, the volume measure $\Phi$ depends only on the first derivatives of $\varphi_a$ and this dependance is linear. The fields $\varphi_a$ do not have their own dynamical equations of motion: they are auxiliary fields. All their dynamical effect is displayed in two ways: 1) in the appearance of the scalar field $\zeta$ and its gradient in all equations of motion; 2) in generating the algebraic constraint (\ref{eq:constraint1}) which determines $\zeta$ as a function of the metric and matter fields.

\subsection{Basic Scalar Field Model}
We shall study a model of the type (\ref{eq:classmodel}) with the matter Lagrangians describing a single scalar field $\phi$. The action now takes the form
\begin{align}\label{eq:singlescalaraction}
&\mathcal{S}=\mathcal{S}_g+\mathcal{S}_\phi, \text{ where,}\nonumber\\
&\mathcal{S}_g=-\frac{1}{\kappa}\int d^4x(\Phi+b_g\sqrt{-g})R(\Gamma,g),\nonumber\\
&\mathcal{S}_{\phi}=\int d^4x\left[(\Phi+b_\phi\sqrt{-g})\frac{1}{2}g^{\mu\nu}\phi,_{\mu}\phi,_{\nu}-\Phi V_1(\phi)-\sqrt{-g}V_2(\phi)\right].
\end{align}

A new dimensionless factor $b_\phi$ appears here for we have no reason to assume a similar coupling of the scalar field to the two volume measures. $V_1(\phi)$ and $V_2(\phi)$ are potential-like functions, we will later see that the actual effective potential that governs the behavior of $\phi$ is a complicated function of both $V_1(\phi)$ and $V_2(\phi)$.

Putting these terms into the constraint (\ref{eq:constraint1}) yields
\be\label{eq:constraintsinglescalar}
(\zeta-b_g)[sM^4+V_1(\phi)]+2V_2(\phi)+b_g\frac{\delta}{2}g^{\alpha\beta}\phi,_{\alpha}\phi,_{\beta}=0,
\ee
where $\delta=(b_g-b_\phi)/b_g$. Since this constraints gives $\zeta\neq0$, the connection (\ref{eq:connection}) is not the Christoffel's connection obtained from $g_{\mu\nu}$, this means that the space-time with the metric $g_{\mu\nu}$ is non-Riemannian. To obtain a physical meaning to the model we must perform a transition to a new metric
\be\label{eq:transformation}
\tilde{g}_{\mu\nu}=(\zeta+b_g){g}_{\mu\nu}.
\ee 

Now, the connection $\Gamma_{\mu\nu}^{\lambda}$ becomes equal to the Christoffel connection $\{_{\mu\nu}^{\lambda} \}$ of the metric $\tilde{g}_{\mu\nu}$. The space-time, equipped with the metric $\tilde{g}_{\mu\nu}$ now becomes a (semi-)Riemannian manifold. This is why we refer to the set of dynamical variables using the metric $\tilde{g}_{\mu\nu}$  as the Einstein frame. One should note that the transformation (\ref{eq:transformation}) is not a conformal transformation since $(\zeta+b_g)$ is sign indefinite. However, $\tilde{g}_{\mu\nu}$ is a regular semi-Riemannian metric. In the Einstein frame, the gravitational equations of motion (\ref{eq:metricvar}) take on the canonical GR form with the same $\kappa=16\pi G$:
\be
G_{\mu\nu}(\tilde{g}_{\alpha\beta})=\frac{\kappa}{2}T_{\mu\nu}^{\text{eff}},
\ee
where $G_{\mu\nu}(\tilde{g}_{\alpha\beta})$ is the Einstein tensor in the Riemannian space-time with the metric $\tilde{g}_{\alpha\beta}$, and the effective stress-energy tensor reads
\begin{equation}\label{eq:est}
T^{\text{eff}}_{\mu\nu}=\frac{\zeta+b_{\phi}}{\zeta+b_{g}}\left({\bold{\phi}},_{\mu}{\bold{\phi}},_{\nu}-\frac{1}{2}{\tilde{g}}_{\mu\nu}{\tilde{g}}^{\alpha\beta}{\bold{\phi}},_{\alpha}{\bold{\phi}},_{\beta}\right)-{\tilde{g}}_{\mu\nu}\frac{b_g-b_{\phi}}{2(\zeta+b_g)}{\tilde{g}}^{\alpha\beta}{\bold{\phi}},_{\alpha}{\bold{\phi}},_{\beta}+{\tilde{g}}_{\mu\nu}V_{\text{eff}}(\zeta,M),
\end{equation}
where the effective potential $V_{\text{eff}}(\zeta,M)$ is calculated using
\begin{equation}\label{eq:Vefscalar}
V_{\text{eff}}(\phi;\zeta,M)=\frac{b_g\left[sM^4+V_1(\phi)\right]-V_2(\phi)}{\left(\zeta+b_g\right)^2}.
\ee

In the Einstein frame the constraint (\ref{eq:constraintsinglescalar}) takes the form
\be
(\zeta-b_g)[sM^4+V_1(\phi)]+2V_2(\phi)+b_g\frac{\delta}{2}\tilde{g}^{\alpha\beta}\phi,_{\alpha}\phi,_{\beta}(\zeta+b_g)=0,
\ee
which determines the structure of the function $\zeta$ which appears in the rest of the equations of motion.

The scalar field equation following from (\ref{eq:singlescalaraction}) reads, in the Einstein frame, as:
\be
\frac{1}{\sqrt{-\tilde{g}}}\partial_{\mu}\left[\frac{\zeta+b_{\phi}}{\zeta+b_g}\sqrt{-\tilde{g}}\tilde{g}^{\mu\nu}\partial_{\nu}\phi\right]+\frac{\zeta V_1'+V_2'}{(\zeta+b_g)^2}=0
\ee

\subsection{Fine Tuning Free Transition To $\Lambda=0$ State}\label{ftft}
One interesting trait of the Two Measure Field Theory is a rather simple and natural transition to a state with zero cosmological constant. This resolution of the old CC problem can be achieved with a model as simple as that discussed in the previous section (\ref{eq:singlescalaraction}). We notice that the $\zeta$ dependence of $V_{\text{eff}}(\phi;\zeta,M)$ is of the form of an inverse square like $\left(\zeta+b_g\right)^{-2}$, this has a key role in the resolution of the problem. Moreover, one can show that even if quantum corrections to the underlying actions generate nonlinear coupling terms between the curvature and the scalar field like $\propto R(\Gamma,g)\phi^2$ in both $L_1$ and $L_2$, the general form of the $\zeta$ dependence in $V_{\text{eff}}(\phi;\zeta,M)$ remains similar $V_{\text{eff}}\propto\left(\zeta+f(\phi)\right)^{-2}$ where $f(\phi)$ is a function of the scalar field alone. The fact that only such dependence emerges and that $\zeta$ is absent from the numerator of $V_{\text{eff}}(\phi;\zeta,M)$ is a direct result of our assumption that the Lagrangians do not depend on the measure fields $\varphi_a$.

Generically in the action (\ref{eq:singlescalaraction}) $b_g\neq b_{\phi}$, this results in a nonlinear kinetic term(a k-essence type dynamics) in the Einstein frame. For the purposes of our discussion however it is enough to take the simplified model with $b_g=b_\phi$. Although this is technically speaking a fine tuning, it is not relevant to the point we are trying to make which holds for the general case as well since the nonlinear kinetic term has no effect on the zero CC problem. This means setting $\delta=0$. Solving the constraint gives
\be\label{eq:zetascalar}
\zeta(\phi)=\frac{b_g[sM^4+V_1(\phi)]-2V_2(\phi)}{[sM^4+V_1(\phi)]},
\ee
and substituting this into the effective potential we get
\be\label{eq:effpotscalar}
V_{\text{eff}}(\phi;M)=\frac{(sM^4+V_1(\phi))^2}{4[b_g(sM^4+V_1(\phi))-V_2(\phi)]}.
\ee

For an arbitrary non-constant $V_1(\phi)$ there exist infinitely many numbers of the integration constant $sM^4$ such that $V_{\text{eff}}(\phi)$ has an absolute minimum at some point $\phi=\phi_0$ where $V_{\text{eff}}(\phi_0)=0$, provided the denominator $4[b_g(sM^4+V_1(\phi))-V_2(\phi)]$ is positive. This affect takes place as $sM^4+V_1(\phi_0)=0$ without fine tuning of the parameters and initial conditions. Note that a choice of a potential in the form of a perfect square in GR is a fine tuning whereas here it emerges naturally from the theory.

To illustrate the point made here let us consider a scalar field model with
\be
V_{1}(\phi)=-\frac{1}{2}\mu_1^2\phi^2\quad, \quad V_{2}(\phi)=V_2^{(0)}-\frac{1}{2}\mu_2^2\phi^2.
\ee 
Recall that adding a constant to $V_{1}$ has no effect on the equations of motion, while $V_2^{(0)}$ absorbs the bare CC and all possible vacuum contributions. We choose a positive integration i.e. $s=1$ and the only restriction on the values of the integration constant $M$ is that the denominator in (\ref{eq:effpotscalar}) remains positive.

Consider spatially flat FRW universe with the metric in the Einstein frame
\be\label{FRWflat}
\tilde{g}_{\mu\nu}=\text{diag}(1,-a^2,-a^2,-a^2)
\ee
where $a=a(t)$ is the scale factor. Each cosmological solution will end up in a $\Lambda=0$ state via damping oscillations of the scalar field $\phi$ towards its absolute minimum $\phi_0$.  From (\ref{eq:zetascalar}) it follows that as $\phi\rightarrow\phi_0$, $|\zeta|\rightarrow \infty$. To be more precise, oscillations of $(sM^4+V_1(\phi))$ around zero are accompanied with singular behavior of the field $\zeta$ whenever $\phi$ crosses $\phi_0$
\be
\frac{1}{\zeta}\sim(sM^4+V_1(\phi))\rightarrow 0\quad\text{as}\quad\phi\rightarrow\phi_0,
\ee
and $\zeta^{-1}$ oscillates around zero with $(sM^4+V_1(\phi))$. Taking into account that the metric in the Einstein frame (\ref{FRWflat}) is regular we deduce by (\ref{eq:transformation}) that the metric $g_{\mu\nu}$ of the underlying action becomes degenerate whenever $\phi=\phi_0$:
\be
g_{00}=\frac{\tilde{g}_{00}}{\zeta+b_g}\sim\frac{1}{\zeta}\rightarrow0,\quad g_{ii}=\frac{\tilde{g}_{ii}}{\zeta+b_g}\sim-\frac{1}{\zeta}\rightarrow0,\quad\text{as }\phi\rightarrow0 ,
\ee
where we have taken into account that the energy density approaches zero and therefore for this cosmological solution the scale factor $a(t)$ remains finite at all times. We see therefore that
\be
\sqrt{-g}\sim\frac{1}{\zeta^2}\rightarrow0\quad\text{and}\quad\Phi=\zeta\sqrt{-g}\sim\frac{1}{\zeta}\rightarrow0\quad\text{as}\quad \phi\rightarrow\phi_{0}
\ee

We recall that the manifold volume measure $\Phi$ is sign indefinite, therefore we are not surprised that it can oscillate about zero and change sign. But TMT shows that including the manifold degrees of freedom into the dynamics of the scalar-gravity system we discover an interesting dynamical effect: the transition to zero CC is accompanied by such oscillations. Similar oscillations simultaneously happen with the components of the metric $g_{\mu\nu}$ used in the underlying action (\ref{eq:singlescalaraction}).

The measure $\Phi$ and the metric $g_{\mu\nu}$ pass zero only in a discrete set of moment in the transition to the $\Lambda=0$ state. Therefore there is no problem with the condition $\Phi\neq0$ used for the solution(\ref{eq:constraint0}). Also, there is no problem with the singularity of $g_{\mu\nu}$ in the underlying action since
\be
\lim_{\phi\rightarrow\phi_0}\Phi g^{\mu\nu}=\text{finite}\quad\text{and}\quad\sqrt{-g}g^{\mu\nu}\sim\frac{1}{\zeta}\rightarrow0\quad\text{as}\quad\phi\rightarrow\phi_0
\ee
The metric in the Einstein frame $\tilde g^{\mu\nu}$ is always regular because degeneracy of $g^{\mu\nu}$ is compensated by singularity of the ratio $\zeta\equiv\Phi/\sqrt{-g}$.

The resolution of the old CC problem is explained in more details in \cite{Guendelman:2006xx} along with an explanation of how TMT avoids Weinbergs no-go theorem \cite{RevModPhys.61.1} which states that in a scalar-tensor field theory one cannot achieve zero CC without fine tuning. Another interesting point is that the zero CC state occurs in TMT in the regime where $\zeta\rightarrow\infty$. From the point of view of TMT this explains why standard theories(with only a metrical volume element $\sqrt{-g}$) cannot resolve the old CC problem without fine tuning. 

\subsection{Gauge Coupling in TMT}\label{gauge}
Let us define what we intend when we call a term 'conformally invariant'. Local conformal invariance is defined in the following way: Let the metric transform as $g_{\mu\nu}\rightarrow\Omega(x)g_{\mu\nu}$ and the measure fields $\varphi_a$ transform according to $\varphi_a\rightarrow\varphi_a'(\varphi(b))$ so that $\Phi\rightarrow\Phi'=J(x)\Phi$ where $J(x)$ is the Jacobian of the transformation of the measure fields $\varphi_a $. This will be a symmetry if $J=\Omega>0$. We shall call any term of the action that is invariant under such a symmetry transformation a 'conformally invariant term'.

We wish to realize a way to implement conformally invariant gauge coupling in the framework of a TMT. This was originally done in an attempt to explain QCD confinement, the fact that free quarks and gluons have never been observed, using TMT \cite{Guendelman2010f}. It has been discovered quite recently that an effective theory which allows both standard Yang-Mills terms of the form $F^a_{\mu\nu}F^{a\mu\nu}$ and terms of the form $\sqrt{F^a_{\mu\nu}F^{a\mu\nu}}$ give rise to confinement. These terms have been coined 'confinement terms'.

We will show that these confinement terms, which seem to be somewhat \textit{ad hoc} in standard field theory, are actually rather natural in TMT. The reason for this is simple: conformally invariant terms in TMT are of two kinds, if they multiply the manifold volume measure $\Phi$ they must have homogeneity 1 with respect to $g^{\mu\nu}$. If they multiply the measure $\sqrt{-g}$ they must have homogeneity 2 with respect to the metric $g^{\mu\nu}$. 

Since $\sqrt{F^a_{\mu\nu}F^{a\mu\nu}}=\sqrt{F^a_{\mu\nu}F^a_{\alpha\beta}g^{\mu\alpha}g^{\nu\beta}}$ then a conformal transformation on such a term yields $\sqrt{F^a_{\mu\nu}F^{a\mu\nu}}\rightarrow\Omega(x)^{-1}\sqrt{F^a_{\mu\nu}F^{a\mu\nu}}$. If there is conformal symmetry then $\Phi\rightarrow\Omega\Phi$ hence the term $\Phi\sqrt{F^a_{\mu\nu}F^{a\mu\nu}}$ is an invariant under conformal transformations of the metric.

Likewise, conformal invariance implies that the term proportional to ${F^a_{\mu\nu}F^a_{\alpha\beta}g^{\mu\alpha}g^{\nu\beta}}$ must come multiplied by the Riemannian measure $\sqrt{-g}$ and then $\sqrt{-g}F^a_{\mu\nu}F^{a\mu\nu}$  is an invariant under conformal transformations of the metric.

We take therefore for the Gauge part of the action the term
\be\label{F}
\mathcal{S}_{F}=-\frac{1}{4}\int d^4x \sqrt{-g}F^{a}_{\alpha\beta}F^{a}_{\mu\nu}g^{\mu\alpha}g^{\nu\beta}-\frac{N}{2}\int d^4x\Phi\sqrt{F^{a}_{\tau\lambda}F^{a}_{\alpha\beta}g^{\tau\alpha}g^{\lambda\beta}}
\ee

\clearpage
\section{'t Hooft-Polyakov-Like Monopole in TMT}\label{sec:thoofttmt}
\subsection{General Calculation of the Energy-Stress Tensor in a Yang-Mills-Higgs theory}
We shall begin our calculation with an underlying action which contains three parts. Following the model (\ref{eq:classmodel}) we will have a regular curvature term. Our matter Lagrangians will now include two contributions, one from a iso-vector scalar field $\phi^a$  and the other from a conformally invariant Gauge term as in (\ref{F})
\begin{align}\label{actionthoofh}
&\mathcal{S}=\mathcal{S}_g+\mathcal{S}_\phi+\mathcal{S}_{F},\\
&\mathcal{S}_g=-\frac{1}{\kappa}\int d^4x(\Phi+b_g\sqrt{-g})R(\Gamma,g),\\
&\mathcal{S}_{\phi}=\int d^4x\left[(\Phi+b_\phi\sqrt{-g})\frac{1}{2}g^{\mu\nu}\phi^a,_{\mu}\phi^a,_{\nu}-\Phi V_1(\phi)-\sqrt{-g}V_2(\phi)\right],\\
&\mathcal{S}_{F}=-\frac{1}{4}\int d^4x \sqrt{-g}F^{a}_{\alpha\beta}F^{a}_{\mu\nu}g^{\mu\alpha}g^{\nu\beta}-\frac{N}{2}\int d^4x\Phi\sqrt{F^{a}_{\tau\lambda}F^{a}_{\alpha\beta}g^{\tau\alpha}g^{\lambda\beta}},
\end{align}
%
where ${\bold{\phi}^{a}}_{;\mu}={\bold{\phi}^{a}},_{\mu}+e\varepsilon^{abc}A_{\mu}^{b}{\bold{\phi}^{c}}$ and $F^{a}_{\mu\nu}=A^{a}_{\nu,\mu}-A^{a}_{\mu,\nu}+e\varepsilon^{abc}A_{\mu}^{b}{A_{\nu}^{c}}$. $A^{a}_{\mu}$ is a vector potential and $V_1$ and $V_2$ are some potential-like functions.

Applying the constraint (\ref{eq:constraint0}) we have
\be\label{eq:varphi}
-\frac{1}{\kappa}R(\Gamma,g)
+\frac{1}{2}g^{\mu\nu}{\bold{\phi}^{a}}_{;\mu}{\bold{\phi}^a}_{;\nu}-\frac{N}{2}\sqrt{g^{\tau\alpha}g^{\lambda\beta}F^{a}_{\tau\lambda}F^{a}_{\alpha\beta}}- V_1=sM^{4},
\ee
where $M$ is a constant of integration with the units of mass. We shall choose $s=1$.

The gravitational equations (\ref{eq:metricvar}) now read
\begin{align}\label{eq:omegamn}
&\left(-\frac{1}{\kappa}R_{\mu\nu}(\Gamma)
+\frac{1}{2}{\bold{\phi}^{a}}_{;\mu}{\bold{\phi}^a}_{;\nu}-\frac{N}{2}\frac{g^{\alpha\beta}F^{a}_{\mu\alpha}F^{a}_{\nu\beta}}{\sqrt{g^{\tau\alpha}g^{\lambda\beta}F^{a}_{\tau\lambda}F^{a}_{\alpha\beta}}}\right)\zeta\\
+&\left(-b_g\frac{1}{\kappa}R_{\mu\nu}(\Gamma)+b_\phi\frac{1}{2} {\bold{\phi}^{a}}_{;\mu}{\bold{\phi}^a}_{;\nu} -\frac{1}{2}g^{\alpha\beta}F^{a}_{\mu\alpha}F^{a}_{\nu\beta}\right)\nonumber\\
-&\frac{1}{2}g_{\mu\nu}\left(-b_g\frac{1}{\kappa}R(\Gamma,g)
+b_{\phi}\frac{1}{2}g^{\mu\nu}{\bold{\phi}^{a}}_{;\mu}{\bold{\phi}^a}_{;\nu}-\frac{1}{4}g^{\mu\alpha}g^{\nu\beta}F^{a}_{\mu\nu}F^{a}_{\mu\nu}-V_2\right)=0,
\end{align}
defining, as before(\ref{eq:zeta}), $\zeta=\frac{\Phi}{\sqrt{-g}}$ to be the ratio of the two measures and therefore a scalar field.


We wish to arrive at a second equation for $R(\Gamma,g)$. This is achieved by contracting the gravitational equations with $g^{\mu\nu}$,
\begin{align}\label{eq:omega}
&-\frac{1}{\kappa}R({\Gamma},g)(\zeta-b_g)+\frac{1}{2}g^{\alpha\beta}\bold{\phi}^{a}_{;\alpha}{\bold{\phi}^a}_{;\beta}(\zeta-b_{\phi})-\frac{N}{2}\zeta\sqrt{g^{\tau\alpha}g^{\lambda\beta}F^{a}_{\tau\lambda}F^{a}_{\alpha\beta}}+2V_2=0.
\end{align}
We see now that we have two different equations containing scalar curvature (\ref{eq:varphi}, \ref{eq:omega}):
\begin{align}
&-\frac{1}{\kappa}R({\Gamma},g)(\zeta-b_g)+\frac{1}{2}g^{\alpha\beta}\bold{\phi}^{a}_{;\alpha}{\bold{\phi}^a}_{;\beta}(\zeta-b_{\phi})-\frac{N}{2}\zeta\sqrt{g^{\tau\alpha}g^{\lambda\beta}F^{a}_{\tau\lambda}F^{a}_{\alpha\beta}}+2V_2=0, \\
&-\frac{1}{\kappa}R(\Gamma,g)
+\frac{1}{2}g^{\mu\nu}{\bold{\phi}^{a}}_{;\mu}{\bold{\phi}^a}_{;\nu}-\frac{N}{2}\sqrt{g^{\tau\alpha}g^{\lambda\beta}F^{a}_{\tau\lambda}F^{a}_{\alpha\beta}}- V_1=M^{4}.
\end{align}

These equations must be agree with each other thus we get a consistency condition which we call the \textit{constraint in the original frame}:
\be\label{eq:constraint original}
\frac{1}{2}g^{\alpha\beta}\bold{\phi}^{a}_{;\alpha}{\bold{\phi}^a}_{;\beta}(b_g-b_{\phi})-\frac{Nb_g}{2}\sqrt{g^{\tau\alpha}g^{\lambda\beta}F^{a}_{\tau\lambda}F^{a}_{\alpha\beta}}+(\zeta-b_g)(M^4+V_1)+2V_2=0.
\ee
 
This is equivalent to using the constraint defined by (\ref{eq:constraint1}). Immediately we see that this equation which governs the behavior of the field $\zeta$ is now dependent also on the gauge fields.
 
 Our main goal in this section is to calculate the energy stress-tensor. Our next step in getting to this goal is to calculate the Einstein tensor that corresponds to the underlying action (\ref{actionthoofh}). From equation (\ref{eq:omegamn}) we can extract the curvature tensor:
\scriptsize 
 \begin{align}
 &\frac{1}{\kappa}R_{\mu\nu}(\Gamma)=\frac{1}{2}\frac{(\zeta+b_{\phi})}{(\zeta+b_{g})}\bold{\phi}^{a}_{;\mu}{\bold{\phi}^a}_{;\nu}-\frac{1}{2(\zeta+b_g)}\left(\frac{N\zeta}{\sqrt{g^{\tau\alpha}g^{\lambda\beta}F^{a}_{\tau\lambda}F^{a}_{\alpha\beta}}}+1\right)g^{\alpha\beta}F^{a}_{\mu\alpha}F^{a}_{\nu\beta}\nonumber\\
 &+\frac{1}{2(\zeta+b_g)}g_{\mu\nu}\left((b_g-b_\phi)\frac{1}{2}g^{\alpha\beta}{\phi}^{a}_{;\alpha}{\bold{\phi}^a}_{;\beta}+\frac{1}{4}g^{\tau\alpha}g^{\lambda\beta}F^{a}_{\tau\lambda}F^{a}_{\alpha\beta}+\frac{Nb_g}{2}\sqrt{g^{\tau\alpha}g^{\lambda\beta}F^{a}_{\tau\lambda}F^{a}_{\alpha\beta}}
 -(b_g(M^{4}+V_1)-V_2)\right).
 \end{align}
 \normalsize
 Next, using simple algebra we can calculate the following rather gruesome expression for the Einstein tensor:
 \footnotesize
 \begin{align}
 \frac{1}{\kappa}G_{\mu\nu}\equiv&\frac{1}{\kappa}\left(R_{\mu\nu}(\Gamma)-\frac{1}{2}g_{\mu\nu}g^{\alpha\beta}R_{\alpha\beta}(\Gamma)\right)\nonumber\\
 =&\frac{1}{2}\frac{(\zeta+b_{\phi})}{(\zeta+b_{g})}\left(\bold{\phi}^{a}_{;\mu}{\bold{\phi}^a}_{;\nu}-\frac{1}{2}g_{\mu\nu}g^{\alpha\beta}\bold{\phi}^{a}_{;\alpha}{\bold{\phi}^a}_{;\beta}\right)-\frac{1}{2}\frac{1}{(\zeta+b_g)}\left(g^{\alpha\beta}F^{a}_{\mu\alpha}F^{a}_{\nu\beta}-\frac{1}{4}g_{\mu\nu}g^{\tau\alpha}g^{\lambda\beta}F^{a}_{\tau\lambda}F^{a}_{\alpha\beta}\right)\nonumber\\
 &-\frac{N}{2}\left(\frac{g^{\alpha\beta}F^{a}_{\mu\alpha}F^{a}_{\nu\beta}}{\sqrt{g^{\tau\alpha}g^{\lambda\beta}F^{a}_{\tau\lambda}F^{a}_{\alpha\beta}}}-\frac{1}{2}g_{\mu\nu}\sqrt{g^{\tau\alpha}g^{\lambda\beta}F^{a}_{\tau\lambda}F^{a}_{\alpha\beta}}\right)+\frac{N b_g}{2(\zeta+b_g)}\frac{g^{\alpha\beta}F^{a}_{\mu\alpha}F^{a}_{\nu\beta}}{\sqrt{g^{\tau\alpha}g^{\lambda\beta}F^{a}_{\tau\lambda}F^{a}_{\alpha\beta}}}\nonumber\\
 &-g_{\mu\nu}\frac{(b_g-b_\phi)}{4(\zeta+b_g)}g^{\alpha\beta}\bold{\phi}^{a}_{;\alpha}{\bold{\phi}^a}_{;\beta}+g_{\mu\nu}\frac{1}{\zeta+b_g}\frac{1}{2}\left(b_g(M^4+V_1)-V_2\right).
 \end{align}
 \normalsize
 To continue further to the calculation of an effective energy-stress tensor we must move to the Einstein frame (\ref{eq:transformation}) where Einstein's field equations $T_{\mu\nu}^{\text{eff}}=\frac{2}{\kappa}\tilde{G}_{\mu\nu}$ apply. We make the transformation $\tilde{g}_{\mu\nu}=(\zeta+b_g){g}_{\mu\nu}$ and can then calculate the energy-stress tensor which gives:
 \footnotesize
 \begin{align} \label{eq:stress}
 T_{\mu\nu}^{\text{eff}}=\frac{2}{\kappa}\tilde{G}_{\mu\nu}=&\frac{(\zeta+b_{\phi})}{(\zeta+b_{g})}\left(\bold{\phi}^{a}_{;\mu}{\bold{\phi}^a}_{;\nu}-\frac{1}{2}\tilde{g}_{\mu\nu}\tilde{g}^{\alpha\beta}\bold{\phi}^{a}_{;\alpha}{\bold{\phi}^a}_{;\beta}\right)-\left(\tilde{g}^{\alpha\beta}F^{a}_{\mu\alpha}F^{a}_{\nu\beta}-\frac{1}{4}\tilde{g}_{\mu\nu}\tilde{g}^{\tau\alpha}\tilde{g}^{\lambda\beta}F^{a}_{\tau\lambda}F^{a}_{\alpha\beta}\right)\nonumber\\
 &-N\left(\frac{\tilde{g}^{\alpha\beta}F^{a}_{\mu\alpha}F^{a}_{\nu\beta}}{\sqrt{\tilde{g}^{\tau\alpha}\tilde{g}^{\lambda\beta}F^{a}_{\tau\lambda}F^{a}_{\alpha\beta}}}-\frac{1}{2}\tilde{g}_{\mu\nu}\sqrt{\tilde{g}^{\tau\alpha}\tilde{g}^{\lambda\beta}F^{a}_{\tau\lambda}F^{a}_{\alpha\beta}}\right)+\frac{N b_g}{(\zeta+b_g)}\frac{\tilde{g}^{\alpha\beta}F^{a}_{\mu\alpha}F^{a}_{\nu\beta}}{\sqrt{\tilde{g}^{\tau\alpha}\tilde{g}^{\lambda\beta}F^{a}_{\tau\lambda}F^{a}_{\alpha\beta}}}\nonumber\\
 &-\tilde{g}_{\mu\nu}\frac{(b_g-b_\phi)}{2(\zeta+b_g)}\tilde{g}^{\alpha\beta}\bold{\phi}^{a}_{;\alpha}{\bold{\phi}^a}_{;\beta}+\tilde{g}_{\mu\nu}V_{\text{eff}},
  \end{align}
  \normalsize
 where as in the simple case of the single scalar field (\ref{eq:Vefscalar}) the potential $V_{\text{eff}}$ is given by
 \be\label{eq:Vef}
 V_{\text{eff}}=\frac{\left(b_g(M^4+V_1)-V_2\right)}{(\zeta+b_g)^2}.
 \ee

 The scalar field $\zeta$ is determined by the constraint (\ref{eq:constraint original}) which in the Einstein frame (\ref{eq:transformation}) takes the form
\be\label{eq:constraint einstein}
(\zeta+b_{g})\left[X\delta b_g-\frac{Nb_g}{2}\sqrt{\tilde{g}^{\tau\alpha}\tilde{g}^{\lambda\beta}F^{a}_{\tau\lambda}F^{a}_{\alpha\beta}}+(M^4+V_1)\right]-2\left[b_g(M^4+V_1)-V_2\right]=0,
\ee
with $X=\frac{1}{2}{\tilde{g}}^{\alpha\beta}{\bold{\phi}}^{a}_{;\alpha}{\bold{\phi}}^{a}_{;\beta}$ and $\delta=\frac{b_g-b_{\phi}}{b_g}$.
This gives the solution for $\zeta$
\begin{equation}\label{eq:zeta}
\zeta=\frac{2\left[b_g(M^4+V_1)-V_2\right]}{\left[X\delta b_g-\frac{Nb_g}{2}\sqrt{\tilde{g}^{\tau\alpha}\tilde{g}^{\lambda\beta}F^{a}_{\tau\lambda}F^{a}_{\alpha\beta}}+(M^4+V_1)\right]}-b_g.
\end{equation}

Placing this term back into the effective potential $V_{\text{eff}}$[\ref{eq:Vef}] we get
\begin{equation}\label{veffective}
 V_{\text{eff}}=\frac{\left[X\delta b_g-\frac{Nb_g}{2}\sqrt{\tilde{g}^{\tau\alpha}\tilde{g}^{\lambda\beta}F^{a}_{\tau\lambda}F^{a}_{\alpha\beta}}+(M^4+V_1)\right]^2}{4\left[b_g(M^4+V_1)-V_2\right]}.
\end{equation}
Finally, for later calculations we shall need the energy-stress tensor in a slightly different form, we will raise one of the indices, using $\tilde{g}^{\mu\nu}$ now as this tensor is only meaningful in the Einstein frame,
\footnotesize
 \begin{align} \label{eq:stressud}
 (T^{\text{eff}})^{\mu}_{\nu}=&\frac{(\zeta+b_{\phi})}{(\zeta+b_{g})}\left(\tilde{g}^{\mu\sigma}\bold{\phi}^{a}_{;\sigma}{\bold{\phi}^a}_{;\nu}-\frac{1}{2}\delta^{\mu}_{\nu}\tilde{g}^{\alpha\beta}\bold{\phi}^{a}_{;\alpha}{\bold{\phi}^a}_{;\beta}\right)-\left(\tilde{g}^{\mu\sigma}\tilde{g}^{\alpha\beta}F^{a}_{\sigma\alpha}F^{a}_{\nu\beta}-\frac{1}{4}\delta^{\mu}_{\nu}\tilde{g}^{\tau\alpha}\tilde{g}^{\lambda\beta}F^{a}_{\tau\lambda}F^{a}_{\alpha\beta}\right)\nonumber\\
 &-N\left(\frac{\tilde{g}^{\mu\sigma}\tilde{g}^{\alpha\beta}F^{a}_{\sigma\alpha}F^{a}_{\nu\beta}}{\sqrt{\tilde{g}^{\tau\alpha}\tilde{g}^{\lambda\beta}F^{a}_{\tau\lambda}F^{a}_{\alpha\beta}}}-\frac{1}{2}\delta^{\mu}_{\nu}\sqrt{\tilde{g}^{\tau\alpha}\tilde{g}^{\lambda\beta}F^{a}_{\tau\lambda}F^{a}_{\alpha\beta}}\right)+\frac{N b_g}{(\zeta+b_g)}\frac{\tilde{g}^{\mu\sigma}\tilde{g}^{\alpha\beta}F^{a}_{\sigma\alpha}F^{a}_{\nu\beta}}{\sqrt{\tilde{g}^{\tau\alpha}\tilde{g}^{\lambda\beta}F^{a}_{\tau\lambda}F^{a}_{\alpha\beta}}}\nonumber\\
 &-\delta^{\mu}_{\nu}\frac{(b_g-b_\phi)}{2(\zeta+b_g)}\tilde{g}^{\alpha\beta}\bold{\phi}^{a}_{;\alpha}{\bold{\phi}^a}_{;\beta}+\delta^{\mu}_{\nu}V_{\text{eff}}.
  \end{align}
  \normalsize
\subsection{Gravitational Field of a t' Hooft-Polyakov-Like Monopole in TMT}\label{thoofttmtgravfield}
We now wish to calculate the gravitational field created by a t' Hooft-Polyakov-Like monopole coupled to gravity in the framework of the Two Measure Field Theory. The reason we are calling this configuration a 't Hooft-Polyakov-Like monopole rather then just a 't Hooft-Polyakov monopole comes from the added confinement term we introduced into our action (\ref{F}), technically the 't Hooft Polyakov monopole does not feature such a term but as we have shown in section (\ref{gauge}) these terms come naturally here. The underlying action of our theory is given by (\ref{actionthoofh}) with the Higgs field now being a triplet of scalar fields $\phi^a$ with $(a=1,2,3)$. As we are interested in solutions far away from the monopole core we use a nonlinear $\sigma-$model and force the field, via a constraint, to reside on the sphere defined by ${\phi}^{a}{\phi}^{a}-\eta^2=0$.
\begin{align} \label{eq:stmtconfphi}
S_{TPL}= &\int  \left[-\frac{1}{\kappa}R(\Gamma,g)
+\frac{1}{2}g^{\mu\nu}{\bold{\phi}^{a}}_{;\mu}{\bold{\phi}^a}_{;\nu}-\frac{N}{2}\sqrt{g^{\tau\alpha}g^{\lambda\beta}F^{a}_{\tau\lambda}F^{a}_{\alpha\beta}}- V_1 \right]\Phi d^{4}x\nonumber\\ 
+& \int  \left[-b_g\frac{1}{\kappa}R(\Gamma,g)
+b_{\phi}\frac{1}{2}g^{\mu\nu}{\bold{\phi}^{a}}_{;\mu}{\bold{\phi}^a}_{;\nu}-\frac{1}{4}g^{\mu\alpha}g^{\nu\beta}F^{a}_{\mu\nu}F^{a}_{\mu\nu}-V_2 \right]\sqrt{-g}d^{4}x\nonumber\\
+&\int\frac{\lambda}{2}({\phi}^{a}{\phi}^{a}-\eta^2)d^4 x
\end{align}
The potential-like functions $V_1$ and $V_2$ produce an effective potential $V^{\text{eff}}$ with a global minimum at $|\phi^a|=\eta$, this enters via the constraint so that in this picture the functions $V_1$ and $V_2$ are actually constants. We've stated earlier that a constant $V_1$ has no influence on the equations of motion, while this is indeed the case we shall still keep it in our calculation for aesthetic reasons. We will however demonstrate explicitly that the constant $V_1$ indeed has no real impact on the model. We see that the constant $V_1$ appears in (\ref{eq:zeta},\ref{veffective}) only in term of the form $M^4+V_1$. This means that it can always be absorbed into the constant of integration $M^4$ so it has no real influence on the equations of motion. 

We look for spherically symmetric static solutions for which, in Schwarzschild coordinates, the scalar and vector fields take the form (\ref{eq:fieldansatz}) and (\ref{eq:gaugeansatz}) respectively along with the asymptotic behavior (\ref{eq:asy}). The spherically symmetric metric has the general form \ref{eq:genmetric}
\be
ds^2=A(r)dt^2-B(r)dr^2-r^2(d\theta^2+\sin^2{\theta}d\varphi^2),
\ee
 Inserting the asymptotic behavior (\ref{eq:asy}) for the fields we obtain the following expression for the gauge field tensor
\be
F^a_{\mu\nu}=\frac{g}{r^4}\left(-2\varepsilon_{\mu\nu a}r^2-2\varepsilon_{\mu a b} r_b r_\nu+2\varepsilon_{\nu a b}r_b r_{\mu}+\varepsilon_{\mu\nu b}r_a r_b\right).
\ee
As stated before, when the Higgs field $\phi^a$ takes its VEV the covariant derivative vanishes. Our constraint demands $\phi^a$ to take its VEV hence the energy-stress tensor (\ref{eq:stressud}) reduces to 
\footnotesize
 \begin{align} \label{eq:tmtthooftstress}
 (T^{\text{eff}})^{\mu}_{\nu}=&-\left(\tilde{g}^{\mu\sigma}\tilde{g}^{\alpha\beta}F^{a}_{\sigma\alpha}F^{a}_{\nu\beta}-\frac{1}{4}\delta^{\mu}_{\nu}\tilde{g}^{\tau\alpha}\tilde{g}^{\lambda\beta}F^{a}_{\tau\lambda}F^{a}_{\alpha\beta}\right)-N\left(\frac{\tilde{g}^{\mu\sigma}\tilde{g}^{\alpha\beta}F^{a}_{\sigma\alpha}F^{a}_{\nu\beta}}{\sqrt{\tilde{g}^{\tau\alpha}\tilde{g}^{\lambda\beta}F^{a}_{\tau\lambda}F^{a}_{\alpha\beta}}}-\frac{1}{2}\delta^{\mu}_{\nu}\sqrt{\tilde{g}^{\tau\alpha}\tilde{g}^{\lambda\beta}F^{a}_{\tau\lambda}F^{a}_{\alpha\beta}}\right)\nonumber\\
 &+\frac{N b_g}{(\zeta+b_g)}\frac{\tilde{g}^{\mu\sigma}\tilde{g}^{\alpha\beta}F^{a}_{\sigma\alpha}F^{a}_{\nu\beta}}{\sqrt{\tilde{g}^{\tau\alpha}\tilde{g}^{\lambda\beta}F^{a}_{\tau\lambda}F^{a}_{\alpha\beta}}}+\delta^{\mu}_{\nu}V_{\text{eff}},
  \end{align}
  \normalsize
where the effective potential is now given by
\begin{equation}\label{eq:veftmtthooft}
 V_{\text{eff}}=\frac{\left(\frac{-Nb_g}{2}\sqrt{\tilde{g}^{\tau\alpha}\tilde{g}^{\lambda\beta}F^{a}_{\tau\lambda}F^{a}_{\alpha\beta}}+(M^4+V_1)\right)^2}{4\left[b_g(M^4+V_1)-V_2\right]},
\end{equation}
and the scalar field $\zeta$ by
\begin{equation}\label{eq:zetatmtthooft}
\zeta=\frac{2\left[b_g(M^4+V_1)-V_2\right]}{\left[\frac{-Nb_g}{2}\sqrt{\tilde{g}^{\tau\alpha}\tilde{g}^{\lambda\beta}F^{a}_{\tau\lambda}F^{a}_{\alpha\beta}}+(M^4+V_1)\right]}-b_g.
\end{equation}
Again we see that all terms in this energy-stress tensor should in principle satisfy the conditions given in (\ref{conditions}). As in the classic case(\ref{eq:schwartzgauge}), the gauge field is given in Schwarzschild coordinates by
 \begin{align}
 F^x_{\theta\varphi}&=-g\sin^2\theta\cos\varphi\quad ,\nonumber\\
 F^y_{\theta\varphi}&=-g\sin^2\theta\sin\varphi\quad ,\nonumber\\
 F^z_{\theta\varphi}&=-g\sin\theta\cos\theta\quad .
 \end{align}
 
 Calculating explicitly the stress-energy tensor (\ref{eq:tmtthooftstress}) we obtain
 \footnotesize
  \begin{align}
T^\varphi_\varphi&=T^\theta_\theta=-\frac{g^2}{r^4}\left(\frac{1}{2}+\frac{N^2b_g^2}{8(b_g(M^4+V_1)-V_2)}\right)+\frac{(M^4+V_1)^2}{4\left(b_g(M^4+V_1)-V_2\right)},\\
T^0_0&=T^r_r=\frac{g^2}{r^4}\left[\frac{1}{2}+\frac{N^2b_g^2}{8\left(b_g(M^4+V_1)-V_2\right)}\right]+\frac{g}{r^2}\frac{N}{\sqrt{2}}\left[\frac{b_g(M^4+V_1)-2V_2}{2\left(b_g(M^4+V_1)-V_2\right)}\right]+\frac{(M^4+V_1)^2}{4\left(b_g(M^4+V_1)-V_2\right)}\nonumber.
 \end{align}
 \normalsize
 We see that again, the conditions (\ref{conditions}) indeed hold for this case, This means that  the metric takes the form (\ref{eq:friendly}):
\be
ds^2=A(r)dt^2-A(r)^{-1}dr^2-r^2(d\theta^2+\sin^2{\theta}d\varphi^2).
\ee
Using the general relation (\ref{Ar}) $A(r)=1-\frac{\kappa}{2r}\int_{0}^{r}T_t^t(r')r'^2dr'$ we can calculate the newtonian potential $A(r)$ exactly. This calculation yields
\begin{equation}\label{newtontmtthooft}
A(r)=1-h-\frac{r_s}{r}+\frac{ Q_g^2 }{r^2}-\frac{ \Lambda r^2}{3},
\end{equation}
where the new coefficients are defined by
\begin{align}\label{thooftcoeff}
Q_g^2&=\frac{\kappa g^2}{2}\left(\frac{4\left(b_g(M^4+V_1)-V_2\right)+N^2b_g^2}{8\left(b_g(M^4+V_1)-V_2\right)}\right),\nonumber\\
h&=\frac{\kappa g N }{\sqrt{2}}\left(\frac{b_g(M^4+V_1)-2V_2}{4\left(b_g(M^4+V_1)-V_2\right)}\right),\nonumber\\
\Lambda&=\frac{\kappa(M^4+V_1)^2}{8\left(b_g(M^4+V_1)-V_2\right)},\nonumber\\
r_s&=\frac{M_{\text{c}}\kappa}{2}.
\end{align}
It is interesting to note that this solution encompasses characteristics of both 't Hooft-Polyakov (\ref{thooftmetric}) and global classical monopoles(\ref{globalmetric}). 

\clearpage
\section{Global Monopole in TMT}\label{sec:globaltmt}
\subsection{General calculation of the Energy-Stress Tensor in a Scalar Field iso-vector theory}
We wish to study the behavior of iso-vector scalar fields in the framework of Two Measure Field Theory. Our underlying action will now contain only two parts, one is that of curvature terms and one from a iso-vector scalar field $\phi^a$.  
\begin{align}\label{eq:globalaction}
&\mathcal{S}=\mathcal{S}_g+\mathcal{S}_\phi,\nonumber\\
&\mathcal{S}_g=-\frac{1}{\kappa}\int d^4x(\Phi+b_g\sqrt{-g})R(\Gamma,g),\nonumber\\
&\mathcal{S}_{\phi}=\int d^4x\left[(\Phi+b_\phi\sqrt{-g})\frac{1}{2}g^{\mu\nu}\phi^a,_{\mu}\phi^a,_{\nu}-\Phi V_1(\phi)-\sqrt{-g}V_2(\phi)\right],
\end{align}
where $V_1$ and $V_2$ are some potential-like functions.
Applying the constraint (\ref{eq:constraint0}) we have
\be\label{eq:varphiscalar}
-\frac{1}{\kappa}R(\Gamma,g)
+\frac{1}{2}g^{\mu\nu}{\bold{\phi}^{a}}_{,\mu}{\bold{\phi}^a}_{,\nu}- V_1=sM^{4},
\ee
where again $M$ is a constant of integration with units of mass and we shall once more choose $s=1$. 

The gravitational equations (\ref{eq:metricvar}) now read
\begin{align}\label{eq:omegamnscalar}
&\left(-\frac{1}{\kappa}R_{\mu\nu}(\Gamma)
+\frac{1}{2}{\bold{\phi}^{a}}_{,\mu}{\bold{\phi}^a}_{,\nu}\right)\zeta+\left(-b_g\frac{1}{\kappa}R_{\mu\nu}(\Gamma)+b_\phi\frac{1}{2} {\bold{\phi}^{a}}_{,\mu}{\bold{\phi}^a}_{,\nu}\right)\nonumber\\
&-\frac{1}{2}g_{\mu\nu}\left(-b_g\frac{1}{\kappa}R(\Gamma,g)
+b_{\phi}\frac{1}{2}g^{\mu\nu}{\bold{\phi}^{a}}_{,\mu}{\bold{\phi}^a}_{,\nu}-V_2\right)=0,
\end{align}
defining, as before(\ref{eq:zeta}), $\zeta=\frac{\Phi}{\sqrt{-g}}$ to be the ratio of the two measures and therefore a scalar field. Continuing our procedure, as before we now contract the gravitational equation using $g^{\mu\nu}$ and get
\begin{align}\label{eq:omegascalar}
&-\frac{1}{\kappa}R({\Gamma},g)(\zeta-b_g)+\frac{1}{2}g^{\alpha\beta}\bold{\phi}^{a}_{,\alpha}{\bold{\phi}^a}_{,\beta}(\zeta-b_{\phi})+2V_2=0.
\end{align}
We will have to reconcile the two equations(\ref{eq:omegascalar},\ref{eq:varphiscalar}) containing the curvature $R({\Gamma},g)$, we get that for the iso-vector scalar field the constraint in the original frame takes the form
\be\label{eq:constorigscalar}
\frac{1}{2}g^{\alpha\beta}\bold{\phi}^{a}_{,\alpha}{\bold{\phi}^a}_{,\beta}(b_g-b_{\phi})+(\zeta-b_g)(M^4+V_1)+2V_2=0.
\ee
This constraint alone will govern the behavior of the scalar field $\zeta$. We proceed in our calculation of the energy-stress tensor by calculating the Einstein tensor that corresponds to the underlying action(\ref{eq:globalaction}). From equation (\ref{eq:omegamnscalar}) we extract the Ricci tensor:
 \small
 \begin{align}
 \frac{1}{\kappa}R_{\mu\nu}(\Gamma)=&\frac{1}{2}\frac{(\zeta+b_{\phi})}{(\zeta+b_{g})}\bold{\phi}^{a}_{,\mu}{\bold{\phi}^a}_{,\nu}+\frac{1}{2(\zeta+b_g)}g_{\mu\nu}\left((b_g-b_\phi)\frac{1}{2}g^{\alpha\beta}{\phi}^{a}_{,\alpha}{\bold{\phi}^a}_{,\beta}-(b_g(M^{4}+V_1)-V_2)\right).
 \end{align}
 \normalsize
 The Einstein tensor would be calculated using simple algebra for the result
 
  \begin{align}
 \frac{1}{\kappa}G_{\mu\nu}\equiv&\frac{1}{\kappa}\left(R_{\mu\nu}(\Gamma)-\frac{1}{2}g_{\mu\nu}g^{\alpha\beta}R_{\alpha\beta}(\Gamma)\right)\nonumber\\
 =&\frac{1}{2}\frac{(\zeta+b_{\phi})}{(\zeta+b_{g})}\left(\bold{\phi}^{a}_{,\mu}{\bold{\phi}^a}_{,\nu}-\frac{1}{2}g_{\mu\nu}g^{\alpha\beta}\bold{\phi}^{a}_{,\alpha}{\bold{\phi}^a}_{,\beta}\right)-g_{\mu\nu}\frac{(b_g-b_\phi)}{4(\zeta+b_g)}g^{\alpha\beta}\bold{\phi}^{a}_{,\alpha}{\bold{\phi}^a}_{,\beta}\nonumber\\&+g_{\mu\nu}\frac{1}{\zeta+b_g}\frac{1}{2}\left(b_g(M^4+V_1)-V_2\right).
 \end{align}
 
 To continue further we must make a transition to a frame where the Einstein field equations $T_{\mu\nu}^{\text{eff}}=\frac{2}{\kappa}\tilde{G}_{\mu\nu}$ hold. We make the transformation $\tilde{g}_{\mu\nu}=(\zeta+b_g){g}_{\mu\nu}$ and can then calculate the energy-stress tensor which gives:
 \begin{align} \label{eq:stresscalar}
 T_{\mu\nu}^{\text{eff}}=\frac{2}{\kappa}\tilde{G}_{\mu\nu}=&\frac{(\zeta+b_{\phi})}{(\zeta+b_{g})}\left(\bold{\phi}^{a}_{,\mu}{\bold{\phi}^a}_{,\nu}-\frac{1}{2}\tilde{g}_{\mu\nu}\tilde{g}^{\alpha\beta}\bold{\phi}^{a}_{,\alpha}{\bold{\phi}^a}_{,\beta}\right)-\tilde{g}_{\mu\nu}\frac{(b_g-b_\phi)}{2(\zeta+b_g)}\tilde{g}^{\alpha\beta}\bold{\phi}^{a}_{,\alpha}{\bold{\phi}^a}_{,\beta}+\tilde{g}_{\mu\nu}V_{\text{eff}}.
  \end{align}
   Where as in the simple case of the single scalar field (\ref{eq:Vefscalar}) the potential $V_{\text{eff}}$ is given by
 \be\label{eq:Vefscalar2}
 V_{\text{eff}}=\frac{\left(b_g(M^4+V_1)-V_2\right)}{(\zeta+b_g)^2}.
 \ee
 The scalar field $\zeta$ is calculated using the constraint (\ref{eq:constorigscalar}) which in the Einstein frame defined by equation (\ref{eq:transformation}) has the form
 \begin{equation}\label{eq:consscalareinst}
(\zeta-b_g)\left[M^{4}+V_1\right]+2V_2+\delta b_g(\zeta+b_g)X=0
\end{equation}
where $X=\frac{1}{2}{\tilde{g}}^{\alpha\beta}{\bold{\phi}^a},_{\alpha}{\bold{\phi}^a},_{\beta}$ and $\delta=\frac{b_g-b_{\phi}}{b_g}$.
This gives us the value of the scalar field $\zeta$:
\begin{equation}\label{eq:zetascalar}
\zeta=b_g\frac{M^4+V_1-\delta b_g X}{M^4+V_1+\delta b_g X}-\frac{2V_2}{M^4+V_1+\delta b_g X}.
\end{equation}
Introducing this term into the effective potential $V_{\text{eff}}$ from equation (\ref{eq:Vefscalar2}) we have
\begin{equation}
 V_{\text{eff}}=\frac{\left[X\delta b_g+(M^4+V_1)\right]^2}{4\left[b_g(M^4+V_1)-V_2\right]}.
\end{equation}
 Again, for the sake of future reference we give the energy stress tensor where one of the indices has been raised using $\tilde{g}_{\mu\nu}$,
 \begin{align} \label{eq:stressudscalar}
 (T^{\text{eff}})^{\mu}_{\nu}=&\frac{(\zeta+b_{\phi})}{(\zeta+b_{g})}\left(\tilde{g}^{\mu\sigma}\bold{\phi}^{a}_{,\sigma}{\bold{\phi}^a}_{,\nu}-\frac{1}{2}\delta^{\mu}_{\nu}\tilde{g}^{\alpha\beta}\bold{\phi}^{a}_{,\alpha}{\bold{\phi}^a}_{,\beta}\right)-\delta^{\mu}_{\nu}\frac{(b_g-b_\phi)}{2(\zeta+b_g)}\tilde{g}^{\alpha\beta}\bold{\phi}^{a}_{,\alpha}{\bold{\phi}^a}_{,\beta}+\delta^{\mu}_{\nu}V_{\text{eff}}.
  \end{align}

\subsection{Gravitational Field of a Global Monopole in TMT}\label{globaltmtgravfield}
We now turn our attention to calculating the gravitational field created by a global monopole coupled to gravity in the framework of Two Measure Field Theory. Our underlying action is given by equation (\ref{eq:globalaction}) with the iso-vector Higgs field being a triplet of scalar fields  $\phi_a$ with $(a=1,2,3)$. Like in the previous case, we are interested in solutions far away from the monopole core. We therefore use a nonlinear $\sigma-$model and force the field, via a constraint, to reside on the sphere defined by ${\phi}^{a}{\phi}^{a}-\eta^2=0$.
\begin{align} \label{eq:stmtglobal}
S_{\text{global}}= &\int  \left[-\frac{1}{\kappa}R(\Gamma,g)
+\frac{1}{2}g^{\mu\nu}{\bold{\phi}^{a}},_{\mu}{\bold{\phi}^a},_{\nu}- V_1 \right]\Phi d^{4}x\nonumber\\ 
+& \int  \left[-b_g\frac{1}{\kappa}R(\Gamma,g)
+b_{\phi}\frac{1}{2}g^{\mu\nu}{\bold{\phi}^{a}},_{\mu}{\bold{\phi}^a},_{\nu}-V_2 \right]\sqrt{-g}d^{4}x+\int\frac{\lambda}{2}({\phi}^{a}{\phi}^{a}-\eta^2)d^4 x
\end{align}
As in the previous case, due to the nonlinear $\sigma$-model, the potential-like functions $V_1$ and $V_2$ are now constants. The constant $V_1$ will be shown to have no effect on the gravitational field.Yet again we see that all terms in (\ref{eq:Vefscalar2}, \ref{eq:zetascalar}) containing the constant $V_1$ come as part of the term $M^4+V_1$, and since $M^4$ is a constant of integration we see again that $V_1$ has no physical consequences. 

We look for spherically symmetric static solutions for which, in Schwarzschild coordinates, the iso-vector field take the form (\ref{eq:fieldansatz}) with the constraint forcing the asymptotic behavior $h=1$ i.e. $\phi^a=\eta x^a/r$. The spherically symmetric metric has the general form (\ref{eq:genmetric})
\be
ds^2=A(r)dt^2-B(r)dr^2-r^2(d\theta^2+\sin^2{\theta}d\varphi^2).
\ee
An explicit calculation of the energy-stress tensor (\ref{eq:stressudscalar}) with the field (\ref{eq:fieldansatz}) yields
\footnotesize
\begin{equation}\label{eq:T}
T_\theta^{\theta}=T_\varphi^{\varphi}=-\frac{\eta^4}{r^4}\cdot\frac{(b_g-b_{\phi})^2}{4[b_g(M^4+V_1)-V_2]}+\frac{(M^4+V_1)^2}{4[b_g(M^4+V_1)-V_2]}.
\end{equation}
\begin{equation}\label{eq:T0}
T_0^0=T_r^r=\frac{\eta^4}{r^4}\left(\frac{(b_g-b_{\phi})^2}{4[b_g(M^4+V_1)-V_2]}\right)+\frac{\eta^2}{r^2}\left(\frac{(b_g+b_{\phi})(M^4+V_1)-2V_2}{2[b_g(M^4+V_1)-V_2]}\right)+\left(\frac{(M^4+V_1)^2}{4[b_g(M^4+V_1)-V_2]}\right).
\end{equation}
\normalsize
We see that again, the conditions (\ref{conditions}) indeed hold for this case, This means that  the metric takes the form (\ref{eq:friendly}):
\be
ds^2=A(r)dt^2-A(r)^{-1}dr^2-r^2(d\theta^2+\sin^2{\theta}d\varphi^2).
\ee
Using the general relation (\ref{Ar}) $A(r)=1-\frac{\kappa}{2r}\int_{0}^{r}T_t^t(r')r'^2dr'$ we can calculate the newtonian potential $A(r)$ exactly. This calculation gives:
\begin{equation}\label{newtontmtglobal}
A(r)=1-h-\frac{r_s}{r}+\frac{ Q_g^2 }{r^2}-\frac{ \Lambda r^2}{3},
\end{equation}
where the new coefficients are defined by
\begin{align}\label{globalcoeff}
Q_g^2&=\frac{\kappa \eta^4(b_g-b_{\phi})^2}{8[b_g(M^4+V_1)-V_2]},\nonumber\\
h&=\kappa \eta^2\left(\frac{(b_g+b_{\phi})(M^4+V_1)-2V_2}{4[b_g(M^4+V_1)-V_2]}\right),\nonumber\\
\Lambda&=\frac{\kappa(M^4+V_1)^2}{8[b_g(M^4+V_1)-V_2]},\nonumber\\
r_s&=\frac{M_{\text{c}}\kappa}{2}.
\end{align}
Once again it is quite surprising that this results has a structure that looks like a combination of both 't Hooft-Polyakov (\ref{thooftmetric}) and Global monopoles (\ref{globalmetric}).

\clearpage

\section{Bubble Dynamics in a TMT Monopole Surroundings}\label{cfgvbhj}
As we have demonstrated in the above calculations (\ref{thoofttmtgravfield}) and (\ref{globaltmtgravfield}), the gravitational fields of 't Hooft-Polyakov-like monopoles and of global monopoles take on a similar structure when working in TMT (\ref{newtontmtthooft},\ref{newtontmtglobal}). Let us examine the behavior of a vacuum bubble circumscribed by such a solution. The Newtonian potentials take on the form
\be
A_\text{i}(r)=1,
\ee
and
\be
A_\text{o}(r)=1-h-\frac{2M}{r}+\frac{ Q_g^2 }{r^2}-{ \chi^2 r^2},
\ee
where we have renamed the cosmological constant $\chi^2\equiv\Lambda/3$.$M$ is again represents the mass of the shell and the monopole in natural units. The shell's effective potential now becomes
\footnotesize
\be
V_{\text{eff}}=1-h-\frac{2 M}{r}+\frac{Q^2_g}{r^2}-r^2 \chi^2 -\frac{\left(h+\frac{2 M}{r}-\frac{Q_g^2}{r^2}-K^2 r^2+r^2 \chi^2 \right)^2}{4 K^2 r^2}.
\ee
\normalsize
The equation $V_{\text{eff}}=0$ is of order eight so when looking for its roots we can have at most four positive real roots.  If we set $h=0$ we reduce the problem, up to a cosmological constant which, if taken to be small, is insignificant, to the one previously discussed in section (\ref{bubbletmtthooft}) and likewise if we set $Q_g=0$ we get the possibilities discussed in (\ref{bubbletmtglobal}). If however we take both $h$ and $Q_g$ non-zero, we find other possibilities.

For small enough, but finite, cosmological constants $\chi^2$ and shell surface tensions $K$ we find \textit{breathing solutions} in the case of \textbf{very small, and even zero mass} $M$ we again have \textit{breathing solutions}. As shown in figure (\ref{Fig:small h qsmallcc}). A large cosmological constant or shell tension abolishes these solutions entirely, as shown in figure (\ref{Fig:smallhQ with cc}).

\begin{figure}[h]
\begin{center}
\includegraphics[width=3.3in]{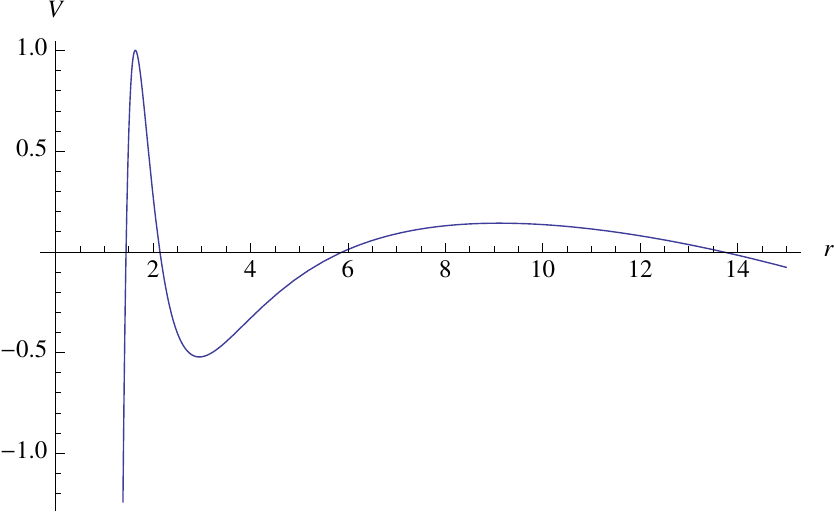}

\caption{\label{Fig:small h qsmallcc} 
The effective bubble potential in a TMT monopole gravitational surroundings, For a monopole strength in the range $0<h<1$, a small monopole charge $1\gg Q_g>0$, a small non-zero cosmological constant $\chi^2>0$ and some positive surface tension. We see that classically the shell has three allowed regions. For small radii the shell will eventually collapse into itself, for large ones the shell will expand indefinitely. However, we see here another classically possible solution. If the initial radius is somewhere between two values $r_{\text{max}}>r>r_{\text{min}}$ the shell will pulsate back and forth between those two values.}
\end{center}
\end{figure}

\begin{figure}[h]
\begin{center}
\includegraphics[width=3.3in]{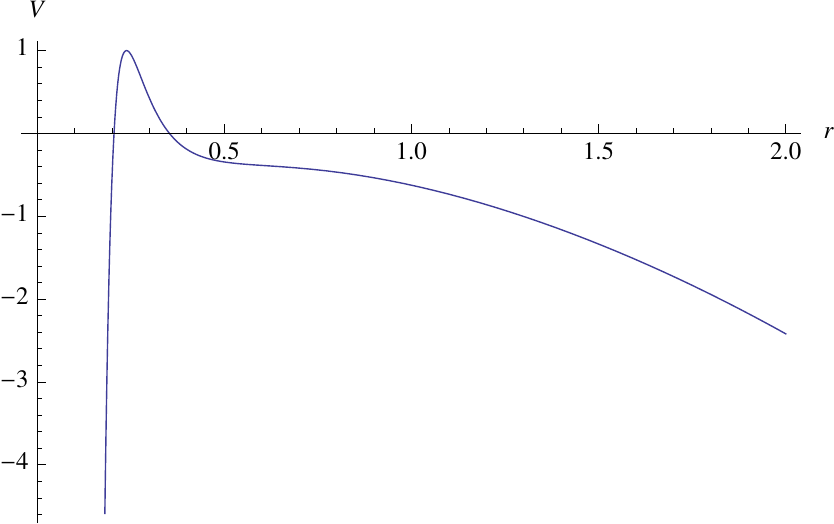}

\caption{\label{Fig:smallhQ with cc}The effective bubble potential in a TMT monopole gravitational surroundings, For a monopole strength in the range $0<h<1$, a small monopole charge $Q_g>0$, a larger cosmological constant $\chi^2$ and some positive surface tension. We see that the increased cosmological constant abolishes the interesting behavior depicted in figure (\ref{Fig:small h qsmallcc}).}
\end{center}
\end{figure}
\clearpage
\section{Discussion and Conclusions}

\subsection{Agreement with Classical Results}
	The first thing we must check is whether the results agrees with the standard result when taking the classical limit. The classical limit in the t Hooft-Polyakov-like monopole case (\ref{thooftcoeff}) corresponds to taking $N=0$ and $M^4+V_1=0$. This gives 
\begin{align}
Q_g^2&=\frac{\kappa g^2}{4},\nonumber\\
h&=0,\nonumber\\
\Lambda&=0,\nonumber\\
r_s&=\frac{M_{\text{c}}\kappa}{2}.
\end{align}	
 Clearly, we see that our result recreates the classical result (\ref{classical thooft field}) and agrees with earlier work\cite{Bais1975,Cho1975}. 
	 
	For the global monopole the classical limit of (\ref{globalcoeff}) corresponds to taking $b_g=b_{\phi}$ and $M^4+V_1=0$. Doing so, we get
\begin{align}
Q_g^2&=0,\nonumber\\
h&=\frac{\kappa \eta^2}{2},\nonumber\\
\Lambda&=0,\nonumber\\
r_s&=\frac{M_{\text{c}}\kappa}{2}.
\end{align}
	and again we see this is in agreement with the classical result (\ref{classical global field}) and is thus compatible with earlier work\cite{Barriola1989, Guendelman1996c}.
	
\subsection{Amplification of Monopole Effects}
	We wish to inspect some implications of the results. Unlike the classical theory where any effect would only be significant for a monopole with strength in the Planck scale, here we see an interesting phenomena. Given a correct set of potentials the influence of the monopole on the gravitational field can be amplified. More precisely, if we set $V_2\approx b_g(M^4+V_1)$ the monopole effect can be immensely amplified. We call this effect \textbf{\textit{Resonant Amplification}} due to obvious reasons. Another possibility for amplification arises once we examine the dependance on the coupling constants. Indeed if we have $b_{\phi}\gg b_{g}$ for the global monopole case or $N\gg0$ for the 't Hooft-Polyakov-like case the effect can also be amplified. 
	
	For a general case we can examine the result and see that for small radii the metric is essentially a Reissner-Nordstr$\ddot{\text{o}}$m metric while for very large radii the metric resembles a de-Sitter space with a cosmological constant. We have chosen our notation in (\ref{classical thooft field}) and (\ref{classical global field}) for this to be apparent.

\subsection{Dealing with Large Distance Divergence}
There is yet another small issue we must attend to. Our solution for the monopole field is in both cases(global and local) asymptotically non-flat. This raises a problem of disagreement of infiniteness of energy. We offer two possible ways to resolve this problem.

One possibility, in the light of work done in global monopole studies, is to apply an explicit cutoff to the energy calculation
\be
M_m=4\pi\int_0^{R}T_t^tr^2dr.
\ee
This cutoff radius $R$ is interpreted as the distance of the nearest anti-monopole. There, the effects of the monopole is canceled by the effect of the anti-monopole and the problem is averted. 

The other possibility stems from the study of Kawai and Matsuo\cite{Kawai2010}. The technique is quite simple, in the light of our discussion of bubble dynamics we postulate a massless charged shell which gives the outermost limit for the effect of the monopole and outside of which by Birkhoff's theorem only a Reissner-Nordstr$\ddot{\text{o}}$m space remains. A solution of equation \ref{hgvhg} in the massless case gives the simple condition $A_{\text{o}}(r)-A_{\text{i}}(r)=0$ which implies a static solution for the radius of the shell. We introduce for the newtonian potentials the relevant functions. For the inner region we take our solution for the monopole field
\begin{equation}
A_{\text{i}}(r)=1-h-\frac{2M}{r}+\frac{ Q_g^2 }{r^2}-{\chi^2 r^2},
\end{equation}
and for the outer region we take the Reissner-Nordstr$\ddot{\text{o}}$m metric with
\be
A_\text{o}(r)=1-\frac{2m}{r}+\frac{q^2}{r^2},
\ee
where $m\geq M$ and $q$ are the combined effective mass and charge ,respectively, of the whole region inside the shell and the shell itself. Equating these two functions we get
\be
\chi^2r^4+hr^2-2(m-M)r+(Q_g^2-q^2)=0.
\ee
The typical form of the LHS function is given in figure (\ref{steadyshell}). The solutions of this equation give us the possible static radii for the massless charged shell.

\begin{figure}[h]
\begin{center}
\includegraphics[width=3.3in]{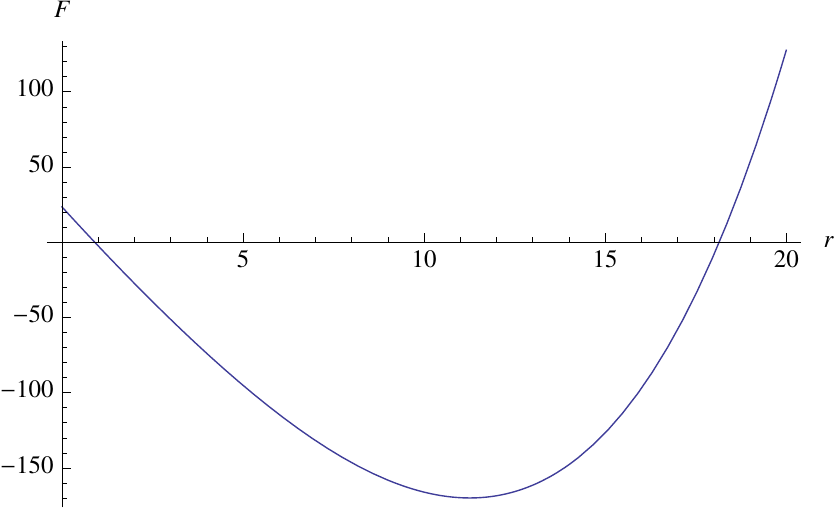}

\caption{\label{steadyshell} 
Typical shape of the function $F(r)=A_{\text{o}}(r)-A_{\text{i}}(r)$. The roots of this function give the possible static radii for a massless charged shell.}
\end{center}
\end{figure}

\subsection{Particle Like States of Bubble Configurations}
As a side note we can look again at figures (\ref{Fig:small h qsmallcc}) and (\ref{steadyshell}). Envisage a situation where the space-time is partitioned into three sections separated by two spherical shells. The innermost section is a vacuum bubble. Around it there is a monopole field. The dynamics of this bubble is similar to that studied in section (\ref{cfgvbhj}). In the third, exterior, region the gravitational field is that of a Reissner-Nordstr$\ddot{\text{o}}$m space-time. This possibility is illustrated in figure (\ref{particle}). This is in a way a realization of the ideas of Kawai and Matsuo\cite{Kawai2010}.

The configuration described above is a classically stable spherically symmetric compact bubble which to an exterior observer can be described by two parameters, $m$ - the mass of the configuration and $q$ - its charge. We interpret this kind of arrangement as classical \textbf{\textit{particle-states}}.

\begin{figure}[h]
\begin{center}
\includegraphics[width=3.3in]{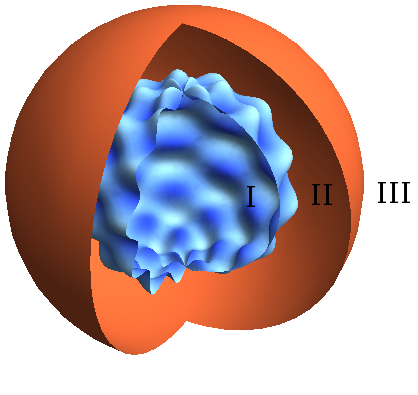}

\caption{\label{particle} 
Illustration of one of the so-called particle-states, with an oscillating interior shell and a static exterior shell. The space-time is partitioned into three sections, I - A vacuum region, II - A region where the gravitational field is that of a monopole, and III - A Reissner-Nordstr$\ddot{\text{o}}$m exterior.}
\end{center}
\end{figure}

\subsection{Universe From a Breathing Bubble}
We've pointed out in sections (\ref{sec:thoofttmt}) and (\ref{sec:globaltmt}) that the potential-like functions $V_1$ and $V_2$ are taken as constants such that the effective potential has absolute minima wherever $\phi^a\phi^a=\eta^2$. However it is possible to imagine more then one configuration of $V_1$ and $V_2$ in which the effective potential might have the same globally minimal value, but the values of $V_1$ and $V_2$  themselves might change. This is essentially a phase transition. On the other hand we've observed in (\ref{thooftcoeff}) and (\ref{globalcoeff}) that $V_1$ is closely related to the cosmological constant and that an increase in $V_1$ is linked to a rise in the cosmological constant. 

Consider a vacuum bubble in the circumscribed by a TMT monopole. As we have demonstrated before there is some region in parameter space where the bubble admits a breathing solution as shown in figure (\ref{Fig:small h qsmallcc}). Now, if by some dynamical process as described in the above paragraph the cosmological constant rises, the bubbles effective potential will transform to the form of figure (\ref{Fig:smallhQ with cc}) and the once stable bubble will grow indefinitely. This gives rise to an evolution of a universe from a breathing bubble. This follows some of the ideas presented by Guendelman and Sakai\cite{Guendelman2008}. Notice that these arguments are all valid on the classical level.

A semi-classical alternative is achieved by a tunneling process. The shell may oscillate for some time in the breathing region, then transport via tunneling to the expanding region and expand indefinitely, creating an empty universe, or to the collapsing region, where it will disappear.

\subsection{Evidence of zero-CC states in a dynamic theory}
We remember that in our model of the monopole the fields were set by a constraint to match the asymptotic field configurations of known theories. We chose this formalism because when working with a dynamical field the calculations can not be done analytically, but only numerically. The constrained model can however provide evidence for some interesting behavior in the dynamical case. 

We look at the form of the effective potential for the t' Hooft-Polyakov-like case where we now allow the potential like functions $V_1$ and $V_2$ to be non-constant and dependent on $\phi$
\begin{equation}\label{eq:veftmtthooft}
 V_{\text{eff}}=\frac{\left(\frac{Nb_g}{2}\sqrt{\tilde{g}^{\tau\alpha}\tilde{g}^{\lambda\beta}F^{a}_{\tau\lambda}F^{a}_{\alpha\beta}}+(M^4+V_1(\phi))\right)^2}{4\left[b_g(M^4+V_1(\phi))-V_2(\phi)\right]}.
\end{equation}
As in the simple case described in section (\ref{ftft}), these exist infinitely many constants of integration $M^4$ such that $V_\text{eff}$ has an absolute minima at some point $\phi=\phi_0$ where $V_{\text{eff}}(\phi_0)=0$. For a monopole configuration this $\phi_0$ will fulfill the condition $|\phi_0|=\eta$. This happens when the following two conditions are satisfied:

\be\label{vjg}
\frac{Nb_g}{2}\sqrt{\tilde{g}^{\tau\alpha}\tilde{g}^{\lambda\beta}F^{a}_{\tau\lambda}F^{a}_{\alpha\beta}}+(M^4+V_1(\phi_0))=0,
\ee
\be\label{asdfg}
b_g(M^4+V_1(\phi_0))-V_2(\phi_0)>0.
\ee

Solving equation (\ref{vjg}) we get 
\be
V_1(\phi_0)=-\frac{Nb_g}{2}\sqrt{\tilde{g}^{\tau\alpha}\tilde{g}^{\lambda\beta}F^{a}_{\tau\lambda}F^{a}_{\alpha\beta}}-M^4=-\frac{Nb_gg}{\sqrt{2}r^2}-M^4,
\ee
which for large radii, in our region of interest away from the core gives
\be\label{V_1away}
V_1(\phi_0)\rightarrow-M^4.
\ee
Combining this result with equation (\ref{asdfg}) we get that away from the core the potential like function $V_2$ must satisfy $V_2<0$.

If we now apply the solution (\ref{V_1away}) we see that far away from the monopole the coefficients (\ref{thooftcoeff}) become
\begin{align}\label{thooftcoefffar}
Q_g^2&=\frac{\kappa g^2}{4}\left(1+\frac{N^2b_g^2}{-4V_2(\phi_0)}\right),\nonumber\\
h&=\frac{\kappa g N }{2\sqrt{2}},\nonumber\\
\Lambda&=0,\nonumber\\
r_s&=\frac{M_{\text{c}}\kappa}{2}.
\end{align}
where $V_2(\phi_0)$ is negative. Notice this solution has zero CC without fine tuning of initial parameters. Further, this solution is independent of the choice of integration constant $M^4$. Also notice $Q_g^2$ is strictly non negative and is zero if and only if the monopole charge $g$ is zero. As for the deficit angle $h$, observe that it is zero when either $g=0$ which accounts to no monopole presence, or for $N=0$ which is the classical limit for this theory.

A similar argument can be made for the global case (\ref{globaltmtgravfield}). This yields coefficients of the form
 \begin{align}\label{globalcoefffar}
Q_g^2&=\frac{\kappa \eta^4(b_g-b_{\phi})^2}{-8V_2},\nonumber\\
h&=\frac{\kappa \eta^2}{2},\nonumber\\
\Lambda&=0,\nonumber\\
r_s&=\frac{M_{\text{c}}\kappa}{2}.
\end{align}
where $V_2$ is negative.  Notice again this solution has zero CC with no fine tuning of initial parameters. Also, this solution is too independent of the choice of integration constant $M^4$. Again notice $Q_g^2$ is strictly non negative and is zero either if the monopole strength $\eta$ is zero, or in the fine tuned case $b_g-b_\phi=0$ which corresponds to the classical limit of this theory. The deficit angle $h$ is zero if and only if the monopole strength itself is zero, i.e. when there is no monopole.

Also notice that as in the simple case described in section (\ref{ftft}) the transition to the zero-CC case will happen via damping oscillations of the effective potentials numerator. These oscillations are again accompanied with singular behavior of the field $\zeta$ whenever $\phi$ is equal to $\phi_0$. This again implies that the underlying metric must be degenerate in the transition, and therefore necessitates the use of a two measure formalism.

\subsection{Outlook on Further Research}
	We will state some of the possible options for further work. 
\begin{itemize}	
\item{In view of the work by Harari et al.\cite{Harari1990d} ,  a numerical calculation of orbits in the field we calculated must be carried out and the question of the existence of bound orbits in a minimally coupled model deserves an answer.}
\item{It would be interesting to investigate the dynamical mechanism which governs the behavior of the potential-like functions $V_1$ and $V_2$ and may therefore lead to bubble destabilization.}
\item{Some studies\cite{Nucamendi2003a}\cite{Nucamendi2001} suggested global monopoles as candidates for causing the gravitational effects known collectively as dark matter. An investigation of our result in that context might yield an interesting outcome.}
\item{In this work we have focused our attention to scalar field monopoles. Following Guendelman and Rabinowitz\cite{Guendelman1991a} it would be interesting to study the behavior of \textit{stringy monopoles} and discover if the two models, which classically emit the same gravitational field, diverge in a TMT framework.}
\item{ Kawai and Matsuo\cite{Kawai2010} recently found a way to use a clever string-membrane hedgehog configuration and build a nonsingular black hole. A study of their model using TMT may yield further interesting results.}
\end{itemize}
\clearpage


%

\footnotesize
\linespread{1}
\bibliographystyle{unsrtnat}
\bibliography{Thesis}
\end{document}